\documentclass[
  shortnames,nojss]{jss}

\usepackage{orcidlink,thumbpdf,lmodern}

\usepackage[utf8]{inputenc}

\author{
Michael J Mahoney~\orcidlink{0000-0003-2402-304X}\\State University of New York College of Environmental Science and Forestry
}
\title{\pkg{waywiser}: Ergonomic Methods for Assessing Spatial Models}

\Plainauthor{Michael J Mahoney}
\Plaintitle{waywiser: Ergonomic Methods for Assessing Spatial Models}

\Abstract{
Assessing predictive models can be challenging. Modelers must navigate a wide array of evaluation methodologies implemented with incompatible interfaces across multiple packages which may give different or even contradictory results, while ensuring that their chosen approach properly estimates the performance of their model when generalizing to new observations. Assessing models fit to spatial data can be particularly difficult, given that model errors may exhibit spatial autocorrelation, model predictions are often aggregated to multiple spatial scales by end users, and models are often tasked with generalizing into spatial regions outside the boundaries of their initial training data.

The \pkg{waywiser} package for the \proglang{R} language attempts to make assessing spatial models easier by providing an ergonomic toolkit for model evaluation tasks, with functions for multiple assessment methodologies sharing a unified interface. Functions from \pkg{waywiser} share standardized argument names and default values, making the user-facing interface simple and easy to learn. These functions are additionally designed to be easy to integrate into a wide variety of modeling workflows, accepting standard classes as inputs and returning size- and type-stable outputs, ensuring that their results are of consistent and predictable data types and dimensions. Additional features make it particularly easy to use \pkg{waywiser} along packages and workflows in the tidymodels ecosystem.
}

\Plainkeywords{spatial modeling, model assessment, applicability domains, R}

\Address{
    Michael J Mahoney\\
    State University of New York College of Environmental Science and Forestry\\
    Graduate Program in Environmental Science\\
1 Forestry Drive\\
Syracuse, NY, USA\\
  E-mail: \email{mjmahone@esf.edu}\\
  URL: \url{https://mm218.dev/}\\~\\
  }

\usepackage{longtable,booktabs,array}
\usepackage{calc} 
\usepackage{etoolbox}
\makeatletter
\patchcmd\longtable{\par}{\if@noskipsec\mbox{}\fi\par}{}{}
\makeatother
\IfFileExists{footnotehyper.sty}{\usepackage{footnotehyper}}{\usepackage{footnote}}
\makesavenoteenv{longtable}

\usepackage{amsmath}

\begin{document}

\hypertarget{introduction}{%
\section{Introduction}\label{introduction}}

Assessing predictive models can be challenging. Modelers must pick from a wide variety of evaluation approaches, each of which may give different and even contradictory results \citep{reich1999}, and ensure that their chosen approach will properly estimate their model's performance when generalizing to observations not used to train the model. Even more complicated is assessing models fit to spatial data, where errors may not be randomly distributed across the study area \citep{legendre1989}, predictions are often aggregated into larger units which may compound existing spatial error patterns, requiring model accuracy assessments at multiple spatial scales \citep{riemann2010}, and models are often used to predict areas outside of the spatial boundary of the initial study area \citep{meyer2021}.

Statistical software can help mitigate some of the complexity created by the array of considerations and approaches for evaluating models fit to spatial data. By providing a common user interface for multiple well-established evaluation procedures, software can make it easier for users to switch between evaluation approaches as appropriate for their current task, helping reduce some of the cognitive load associated with switching between different tasks \citep{roehm2012}. User interfaces should also make it easy to follow scientific and statistical best practices, and similarly make it difficult to commit methodological errors \citep{tidymodels}.

Many excellent \proglang{R} packages aim to help reduce this complexity and promote best practices by addressing individual aspects of model evaluation. Among many others, packages like \pkg{yardstick} \citep{yardstick}, \pkg{metrica} \citep{metrica}, and \pkg{hydroGOF} \citep{hydroGOF} provide suites of metrics for model assessment, providing standard interfaces for calculating model accuracy and agreement given vectors of observed and predicted values. Packages like \pkg{spdep} \citep{spdep} and \pkg{rgeoda} \citep{rgeoda}, meanwhile, provide measures of spatial autocorrelation, helping modelers assess the spatial distribution of model errors. Finally, several packages implement additional evaluation approaches beyond standard error assessments; for instance, \pkg{CAST} \citep{CAST} and \pkg{applicable} \citep{applicable} implement approaches for calculating model applicability domains.

The new \pkg{waywiser} package implements elements of each of these approaches, while providing a consistent, ergonomic user interface for each aspect of model assessment. Functions in \pkg{waywiser} provide new implementations of several popular assessment metrics from the spatial modeling literature, and provide a wrapper around functions for calculating spatial autocorrelation metrics. Additional functions provide an approach for assessing model predictions aggregated to multiple spatial scales, and a new implementation of the dissimilarity index and area of applicability from \citet{meyer2021}. These functions share a consistent interface, with standardized argument names and definitions, making it easy for users to learn how to use the package, and to switch between evaluation approaches as desired.

Outputs from \pkg{waywiser} are both type-stable and size-stable, making \pkg{waywiser} functions both predictable and easy to program with. Functions in \pkg{waywiser} additionally accept inputs and return outputs using standard classes, using objects from the popular \pkg{sf} \citep{sf} package for spatial data and simple data frames and vectors otherwise. This predictability and reliance on well-established classes makes it easy to use \pkg{waywiser} with the majority of modeling tools in \proglang{R}. Additional features make it particularly easy to combine \pkg{waywiser} with packages in the tidymodels modeling framework \citep{tidymodels}. For instance, while \pkg{waywiser} does not itself provide any functions for performing cross-validation or hyperparameter selection, functions from waywiser integrate naturally with the \pkg{tune} package \citep{tune}, allowing for cross-validated model assessment using data splits from \pkg{rsample} \citep{rsample} and \pkg{spatialsample} \citep{spatialsample} and automated hyperparameter selection using \pkg{dials} \citep{dials}.

The rest of this article walks through features in \pkg{waywiser}, starting with functions implementing (or wrapping implementations of) model assessment metrics (Section \ref{sec-assessment-metrics}), followed by methods for assessing model performance when aggregating predictions across multiple spatial scales (Section \ref{sec-multi-scale}), then by functions for calculating the applicability domain of a model (Section \ref{sec-aoa}). An additional section discusses how \pkg{waywiser} integrates with the tidymodels modeling framework (Section \ref{sec-interop}).

\hypertarget{example-data}{%
\section{Example data}\label{example-data}}

For demonstration purposes, this paper will assess a linear model fit to the \texttt{worldclim\_simulation} data included in \pkg{waywiser}, containing a random sample of 10,000 points from the WorldClim Bioclimatic variables data set \citep{worldclim}. Variable ``\texttt{bio2}'' records the mean monthly diurnal temperature range, ``\texttt{bio10}'' the mean temperature of the warmest 3 months of the year, ``\texttt{bio13}'' the precipitation of the wettest month of the year, and ``\texttt{bio19}'' the precipitation of the coldest 3 months of the year. A final variable, ``\texttt{response}'', was simulated using the \pkg{virtualspecies} package following examples in \pkg{CAST} \citep{virtualspecies, CAST}.

To create this model, we will first split our data into training and test sets, resembling a standard predictive modeling workflow. For simplicity, we will assign observations to these sets at random; in actual practice, it would be best to use spatial cross-validation approaches in order to address any spatial autocorrelation in the response variable \citep{spatialsample}.

\begin{CodeChunk}
\begin{CodeInput}
R> set.seed(1107)
R> data("worldclim_simulation", package = "waywiser")
R> worldclim_training <- sample(nrow(worldclim_simulation),
+                              nrow(worldclim_simulation) * 0.8)
R> worldclim_testing <- worldclim_simulation[-worldclim_training, ]
R> worldclim_training <- worldclim_simulation[worldclim_training, ]
\end{CodeInput}
\end{CodeChunk}

We then fit a linear model using base \proglang{R}'s \texttt{lm()} function, and use the resulting model to generate predictions for our test set:

\begin{CodeChunk}
\begin{CodeInput}
R> worldclim_model <- lm(response ~ bio2 + bio10 + bio13 + bio19,
+                       data = worldclim_training)
R> worldclim_testing$predictions <- predict(worldclim_model, 
+                                          worldclim_testing)
\end{CodeInput}
\end{CodeChunk}

\hypertarget{sec-assessment-metrics}{%
\section{Model assessment metrics}\label{sec-assessment-metrics}}

\hypertarget{agreement-metrics}{%
\subsection{Agreement metrics}\label{agreement-metrics}}

Several functions in \pkg{waywiser} revolve around calculating model agreement metrics: numeric indices of how closely model predictions (which we refer to as \(\hat{y}\)) align with another data set (\(y\)), with \(y\) typically (but not necessarily) representing ``true'' measured values. This set of metrics generally originated within the spatial modeling literature and are most popular for assessing models fit to spatial data, but do not incorporate any geographic information into their calculation.

For instance, in a series of papers, Willmott \citetext{\citeyear{willmott1980}; \citeyear{willmott1981}; \citeyear{willmott1982}} introduced a index of agreement, \(d\). This metric represents the agreement of model predictions (\(\hat{y}\)) with observed values (\(y\)) as the ratio of the sum of squared differences to the sum of the absolute values of differences in predicted and observed values from the observed mean (\(\bar{y}\)); that is:

\begin{equation}
d = 1 - \frac{\sum_{i=1}^{n}{\left(\hat{y}_{i}-y_{i}\right)^{2}}}{\sum_{i=1}^{n}{\left(\left|\hat{y}_{i}-\bar{y}\right| + \left|y_{i}-\bar{y}\right|\right)^{2}}}
\label{eq:willmott-d}
\end{equation}

This formulation means that \(d\) is bounded \([0, 1]\), with higher values of \(d\) indicating greater agreement between \(y\) and \(\hat{y}\). As a dimensionless metric, \(d\) is a useful tool for comparing models; however, the use of summed differences in the numerator means that \(d\) is oversensitive to outliers in \(\hat{y}_{i} - y_{i}\) \citep{legates1999}. To address this concern, Willmott et al. \citeyearpar{willmott1985} introduced a revised metric named \(d_1\), using the sum of the absolute values of differences in the place of the sum of squared differences and no longer squaring the denominator:

\begin{equation}
d_1 = 1 - \frac{\sum_{i=1}^{n}{\left|\hat{y}_{i}-y_{i}\right|}}{\sum_{i=1}^{n}{\left(\left|\hat{y}_{i}-\bar{y}\right| + \left|y_{i}-\bar{y}\right|\right)}}
\label{eq:willmott-d1}
\end{equation}

As with \(d\), \(d_1\) is bounded \([0, 1]\) with higher values indicating improved agreement; Willmott \citeyearpar{willmott2011} notes that \(d_1\) approaches 1 more slowly than \(d\), allowing for finer-grained comparisons between well-performing models.

Willmott et al.~revisited these indices twenty-five years later \citeyearpar{willmott2011}, noting that interpretation of \(d\) and \(d_1\) was made difficult both by the limited range of the metric and by the inclusion of \(\hat{y}\) in the denominator, which made the scaling factor of the agreement metric dependent upon the model itself. To address this, they introduce a new metric \(d_r\), such that:

\begin{equation}
d_r = 
\left\{
\begin{array}{l}
1 - \dfrac{\sum_{i=1}^{n}{\left|\hat{y}_{i}-y_{i}\right|}}{c\sum_{i=1}^{n}{\left|y_{i}-\bar{y}\right|}}, \text{ when}\\\\ \sum_{i=1}^{n}{\left|\hat{y}_{i}-y_{i}\right|} \leq c\sum_{i=1}^{n}{\left|y_{i}-\bar{y}\right|}\\\\
\dfrac{c\sum_{i=1}^{n}{\left|y_{i}-\bar{y}\right|}}{\sum_{i=1}^{n}{\left|\hat{y}_{i}-y_{i}\right|}} - 1 \text{ otherwise}
\end{array}
\right\}
\label{eq:willmott-dr}
\end{equation}

Where \(c\) is a scaling constant set to 2. A full derivation is provided in \citet{willmott2011}.

Compared to \(d\) and \(d_1\), \(d_r\) provides a larger metric range (being bounded \([-1, 1]\), with 1 indicating perfect agreement) and is more directly interpretable; \(d_r\) is proportional to the mean absolute error divided by \(c\) times the mean absolute deviation.

These agreement metrics are all asymmetric, assuming that \(y\) values are more accurate than \(\hat{y}\). This makes this set of metrics useful when comparing model predictions against measured values, as measured values used to train models are generally assumed to be more trustworthy than model predictions. However, in some model assessment scenarios it can be desirable to treat both \(y\) and \(\hat{y}\) as capable of containing error. For instance, when comparing two distinct sets of model predictions, it is typically desirable to not treat whichever set of predictions is labeled as \(y\) as being inherently more accurate.

For this reason, \citet{ji2006} introduce an agreement coefficient, \(\operatorname{AC}\), which is symmetrical and allows for errors in both \(y\) and \(\hat{y}\):

\begin{equation}
\operatorname{AC} = 1 - \frac{\sum_{i=1}^{n}{\left(\hat{y}_{i}-y_{i}\right)^{2}}}{\sum_{i=1}^{n}{\left(\left|\bar{\hat{y}}-\bar{y}\right|+\left|\hat{y}_{i}-\bar{\hat{y}}\right|\right)\left(\left|\bar{\hat{y}}-\bar{y}\right|+\left|y_{i}-\bar{y}\right|\right)}}
\label{eq:ac}
\end{equation}

Unlike \(d\) and related metrics, \(\operatorname{AC}\) is symmetrical, producing identical values if \(y\) and \(\hat{y}\) are reversed. This makes \(\operatorname{AC}\) a preferable metric for comparing predictions from models, as it does not assume either set of predictions to be more accurate than the other. \citet{ji2006} describe \(\operatorname{AC}\) as bounded \([0, 1]\), as when \(\sum_{i=1}^{n}{y_i} = \bar{y}\) and \(\sum_{i=1}^{n}{\hat{y}_i} = \bar{\hat{y}}\) (as would occur under a null model) the fractional term simplifies to 1 and thus \(\operatorname{AC}\) is 0. In practice however, the null model is not the true lower bound for how poorly two data sets can agree with one another; it is entirely possible for poor models with large differences between \(y_i\) and \(\hat{y}_i\) to produce negative \(\operatorname{AC}\) estimates, with true bounds of \((-\infty, 1]\). Worse agreement between \(y_i\) and \(y\) produces smaller values.

In addition to these dimensionless agreement metrics, these authors also suggest a host of metrics for model assessment in units of \(y\), which may then be decomposed into systematic and unsystematic components. For instance, \citet{willmott1981} walks through the decomposition of the familiar mean squared error (\(\operatorname{MSE}\)):

\begin{equation}
\operatorname{MSE} = \frac{1}{n}\sum_{i=1}^{n}{\left(\hat{y}_{i}-y_{i}\right)^{2}}
\label{eq:mse}
\end{equation}

Into its systematic and unsystematic components, with the systematic component of \(\operatorname{MSE}\) given by:

\begin{equation}
\operatorname{MSE}_{s} = \frac{1}{n}\sum_{i=1}^{n}{\left(\hat{y}_{i}-y_{i}^\prime{}\right)^{2}}
\label{eq:mse-s}
\end{equation}

And the unsystematic component of \(\operatorname{MSE}\) given by:

\begin{equation}
\operatorname{MSE}_{u} = \frac{1}{n}\sum_{i=1}^{n}{\left(y_{i}-{y}_{i}^\prime{}\right)^{2}}
\label{eq:mse-u}
\end{equation}

Where \(y^\prime{}\) is the predicted value of \(y\) from the linear regression model \(y^\prime{} = a + b\hat{y}\). These two components sum to \(\operatorname{MSE}\):

\begin{equation}
\operatorname{MSE} = \operatorname{MSE}_{s} + \operatorname{MSE}_{u}
\label{eq:mse-decomp}
\end{equation}

As \(\operatorname{MSE}\) is in squared units of \(y\), it is typically more useful to use the root mean squared error (\(\operatorname{RMSE}\)) and its systematic and unsystematic components, calculated by taking the square root of \(\operatorname{MSE}\) and its components.

\citet{ji2006} present a similar decomposition for their \(\operatorname{AC}\) metric, using a geometric mean functional relationship (GMFR) model in place of the linear regression to allow for errors in both \(y\) and \(\hat{y}\) \citep{draper1997}. The GMFR is estimated such that:

\begin{equation}
\begin{aligned}
y^\prime{} &= a + b\hat{y} \\
b &= \pm \left(\frac{\sum_{i=1}^{n}{\left(\hat{y}_{i}-\bar{\hat{y}}\right)^{2}}}{\sum_{i=1}^{n}{\left({y}_{i}-\bar{{y}}\right)^{2}}}\right)^{\frac{1}{2}} \\
a &= \bar{\hat{y}} - b\bar{y}
\end{aligned}
\label{eq:gmfr}
\end{equation}

Where the sign of \(b\) is the same sign as the correlation coefficient between \(y\) and \(\hat{y}\). This regression equation can be reversed to predict \(\hat{y}\), represented by \(\hat{y}^\prime{}\), as a function of \(y\):

\begin{equation}
\hat{y}^\prime{} = -\frac{a}{b} + \frac{1}{by}
\label{eq:gmfr-flip}
\end{equation}

\citet{ji2006} use these quantities to decompose the sum of squared differences into systematic and unsystematic components, which they refer to as the systematic and unsystematic sum of product differences (\(\operatorname{SPD}\)). The unsystematic component of \(\operatorname{SPD}\) is defined as:

\begin{equation}
\operatorname{SPD}_u = \sum_{i=1}^{n}{\left[\left(\left|\hat{y}_{i}-\hat{y}_i^\prime{}\right|\right)\left(\left|{y}_{i}-{y}_i^\prime{}\right|\right)\right]}
\label{eq:spdu}
\end{equation}

While the systematic component is found by subtracting \(\operatorname{SPD}_u\) from the sum of squared differences:

\begin{equation}
\operatorname{SPD}_{s} = \left(\sum_{i=1}^{n}{\left(\hat{y}_{i}-y_{i}\right)^{2}}\right) - \left(\sum_{i=1}^{n}{\left[\left(\left|\hat{y}_{i}-\hat{y}_i^\prime{}\right|\right)\left(\left|{y}_{i}-{y}_i^\prime{}\right|\right)\right]}\right)
\label{eq:spds}
\end{equation}

Taking the arithmetic mean of these terms produces the unsystematic and systematic mean product difference (\(\operatorname{MPD}_{u}\) and \(\operatorname{MPD}_{s}\), respectively):

\begin{equation}
\operatorname{MPD}_{u} = \frac{1}{n}\left(\operatorname{SPD}_u\right)
\label{eq:mpdu}
\end{equation}

\begin{equation}
\operatorname{MPD}_{s} = \frac{1}{n}\left(\operatorname{SPD}_s\right)
\label{eq:mpds}
\end{equation}

These quantities can be expressed as a ratio of MSE to represent the proportion of systematic and unsystematic disagreement between \(y\) and \(\hat{y}\). The square roots of \(\operatorname{MPD}_{u}\) and \(\operatorname{MPD}_{s}\) (\(\operatorname{RMPD}_{u}\) and \(\operatorname{RMPD}_{s}\), respectively) are in units of \(y\) and may be useful ways to describe systematic and unsystematic disagreement in absolute units.

As \(\operatorname{SPD}_u\) and \(\operatorname{SPD}_s\) sum to the sum of squared differences, the numerator of \(\operatorname{AC}\) (Equation \eqref{eq:ac}) can then be decomposed into systematic and unsystematic components by replacing the numerator with the appropriate component of \(\operatorname{SPD}\):

\begin{equation}
\operatorname{AC}_u = 1 - \frac{\operatorname{SPD}_{u}}{\sum_{i=1}^{n}{\left(\left|\bar{\hat{y}}-\bar{y}\right|+\left|\hat{y}_{i}-\bar{\hat{y}}\right|\right)\left(\left|\bar{\hat{y}}-\bar{y}\right|+\left|y_{i}-\bar{y}\right|\right)}}
\label{eq:acu}
\end{equation}

\begin{equation}
\operatorname{AC}_s = 1 - \frac{\operatorname{SPD}_{s}}{\sum_{i=1}^{n}{\left(\left|\bar{\hat{y}}-\bar{y}\right|+\left|\hat{y}_{i}-\bar{\hat{y}}\right|\right)\left(\left|\bar{\hat{y}}-\bar{y}\right|+\left|y_{i}-\bar{y}\right|\right)}}
\label{eq:acs}
\end{equation}

These metrics are all implemented in \pkg{waywiser} using the infrastructure provided by \pkg{yardstick} \citep{yardstick}. Functions are prefixed with \texttt{ww\_}, to help with autocompletion inside of code editors, and all share identical user interfaces. The main version of each metric function takes three primary arguments, namely \texttt{data} (an object inheriting the \texttt{data.frame} S3 class), \texttt{truth} (the name of the column containing \(y\)), and \texttt{estimate} (the name of the column containing \(\hat{y}\)). Both \texttt{truth} and \texttt{estimate} follow tidy evaluation rules, with the main user-noticeable effect being that these arguments accept either quoted or unquoted column identifiers. These functions return a ``tibble'' \citep{tibble} with a single row and three columns, \texttt{.metric} (containing the name of the metric), \texttt{.estimator} (containing the string ``standard'', for compatibility with \pkg{yardstick} and the broader tidymodels ecosystem), and \texttt{.estimate} (containing the metric estimate).

\begin{CodeChunk}
\begin{CodeInput}
R> waywiser::ww_willmott_d(data = worldclim_testing,
+                         truth = response,
+                         estimate = predictions)
\end{CodeInput}
\begin{CodeOutput}
# A tibble: 1 x 3
  .metric    .estimator .estimate
  <chr>      <chr>          <dbl>
1 willmott_d standard       0.919
\end{CodeOutput}
\end{CodeChunk}

When \texttt{data} is a grouped data frame, as produced by the \texttt{group\_by()} function in \pkg{dplyr} \citep{dplyr}, \pkg{waywiser} will calculate metrics independently for each group. In these cases, the resulting tibble will have one row per group and include the columns used to group the data alongside the standard \texttt{.metric}, \texttt{.estimator}, and \texttt{.estimate} columns:

\begin{CodeChunk}
\begin{CodeInput}
R> worldclim_testing$group <- sample(1:2, nrow(worldclim_testing), 
+                                   replace = TRUE)
R> waywiser::ww_willmott_d(data = dplyr::group_by(worldclim_testing, group),
+                         truth = response,
+                         estimate = predictions)
\end{CodeInput}
\begin{CodeOutput}
# A tibble: 2 x 4
  group .metric    .estimator .estimate
  <int> <chr>      <chr>          <dbl>
1     1 willmott_d standard       0.924
2     2 willmott_d standard       0.914
\end{CodeOutput}
\end{CodeChunk}

These functions additionally each have a variant, suffixed with \texttt{\_vec}, which directly accepts numeric vectors to \texttt{truth} and \texttt{estimate}. These functions return a numeric vector with metric estimates.

\begin{CodeChunk}
\begin{CodeInput}
R> waywiser::ww_willmott_d_vec(truth = worldclim_testing$response,
+                             estimate = worldclim_testing$predictions)
\end{CodeInput}
\begin{CodeOutput}
[1] 0.9187938
\end{CodeOutput}
\end{CodeChunk}

Internally, data frame-based functions call their \texttt{\_vec} variants to calculate metrics, ensuring that identical calculations are performed (and therefore, identical results are returned) regardless of which interface is used.

As these functions leverage infrastructure from the \pkg{yardstick} package, the data frame variants can be combined using the \pkg{yardstick} function \texttt{metric\_set()}. This function accepts any number of metric functions as an input and returns a new function to calculate all metrics in a single call. The returned function has the same user interface as the data frame metric functions, accepting the arguments \texttt{data}, \texttt{truth}, and \texttt{estimate} and returning a tibble with the columns \texttt{.metric}, \texttt{.estimator}, and \texttt{.estimate}.

\begin{CodeChunk}
\begin{CodeInput}
R> metrics <- yardstick::metric_set(
+   waywiser::ww_willmott_d, waywiser::ww_willmott_d1,
+   waywiser::ww_willmott_dr, waywiser::ww_systematic_mse,
+   waywiser::ww_unsystematic_mse, waywiser::ww_systematic_rmse,
+   waywiser::ww_unsystematic_rmse, waywiser::ww_agreement_coefficient,
+   waywiser::ww_systematic_agreement_coefficient,
+   waywiser::ww_unsystematic_agreement_coefficient,
+   waywiser::ww_systematic_mpd, waywiser::ww_unsystematic_mpd,
+   waywiser::ww_systematic_rmpd, waywiser::ww_unsystematic_rmpd)
R> print(metrics(data = worldclim_testing,
+               truth = response, estimate = predictions), n = 14)
\end{CodeInput}
\begin{CodeOutput}
# A tibble: 14 x 3
   .metric                            .estimator .estimate
   <chr>                              <chr>          <dbl>
 1 willmott_d                         standard   0.919    
 2 willmott_d1                        standard   0.729    
 3 willmott_dr                        standard   0.759    
 4 systematic_mse                     standard   0.0119   
 5 unsystematic_mse                   standard   0.0000258
 6 systematic_rmse                    standard   0.109    
 7 unsystematic_rmse                  standard   0.00508  
 8 agreement_coefficient              standard   0.658    
 9 systematic_agreement_coefficient   standard   0.980    
10 unsystematic_agreement_coefficient standard   0.677    
11 systematic_mpd                     standard   0.000688 
12 unsystematic_mpd                   standard   0.0112   
13 systematic_rmpd                    standard   0.0262   
14 unsystematic_rmpd                  standard   0.106    
\end{CodeOutput}
\end{CodeChunk}

\hypertarget{autocorrelation-metrics}{%
\subsection{Autocorrelation metrics}\label{autocorrelation-metrics}}

In addition to its set of agreement metrics, \pkg{waywiser} provides a set of functions for measuring spatial autocorrelation in model residuals. These functions provide a thin wrapper over functions provided by the \pkg{spdep} package \citep{spdep, bivand2022, applied2008}, meaning that (unlike the agreement metrics) metric calculations are not implemented directly in \pkg{waywiser}. Equations in this section are largely derived from \citet{spdep}.

Spatial autocorrelation metrics are designed to detect if values among neighboring observations are more related to each other than to a randomly selected observation; that is, if similar values are more clustered together or more dispersed than would be expected at random. In order to assess variable relationships between neighboring observations, it is necessary to first define which observations neighbor one another. A utility function in \pkg{waywiser}, \texttt{ww\_build\_neighbors()}, can be used to do so automatically for classes from the \pkg{sf} package, though it is often preferable for users to more thoughtfully calculate neighbors and provide the resulting object to functions instead. When working with polygon geometries, \texttt{ww\_build\_neighbors()} uses the default behavior of the \texttt{poly2nb()} function from \pkg{spdep}, defining neighbors as any polygons sharing at least one boundary point. We can visualize this using the standard ``moral statistics'' data set from \citet{guerry}:

\begin{CodeChunk}
\begin{CodeInput}
R> data("guerry", package = "waywiser")
R> plot(sf::st_geometry(guerry))
R> plot(waywiser::ww_build_neighbors(guerry), 
+      sf::st_geometry(guerry), add = TRUE)
\end{CodeInput}
\begin{figure}

{\centering \includegraphics{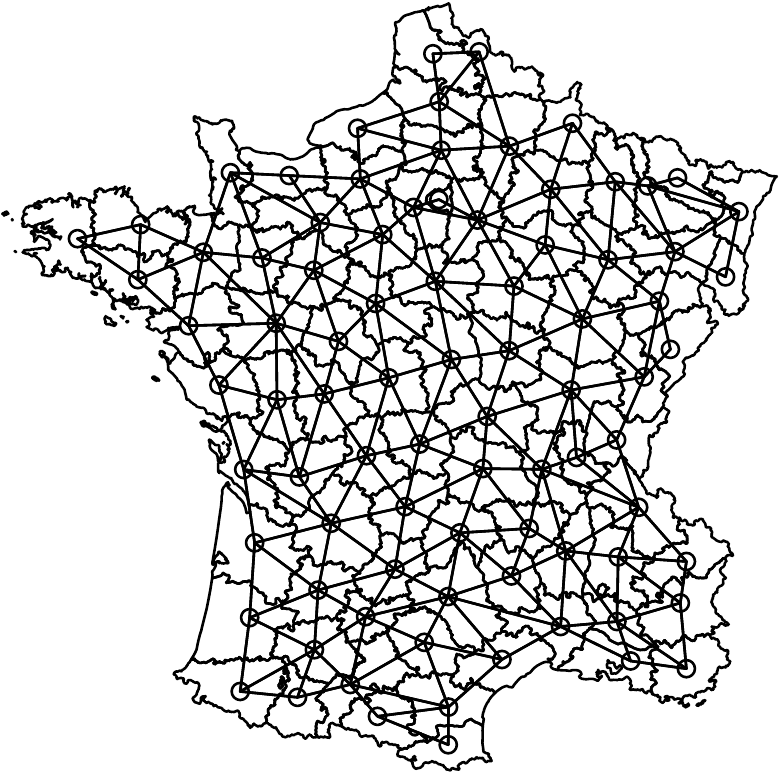} 

}

\caption[Automatically calculated spatial neighbors for departments of France]{Automatically calculated spatial neighbors for departments of France. Lines between department centroids indicate a neighbor relationship.}\label{fig:neighbors-poly}
\end{figure}
\end{CodeChunk}

When working with point geometries, \pkg{waywiser} instead uses the \texttt{knearneigh()} and \texttt{knn2nb()} functions from \pkg{spdep} with \texttt{k\ =\ 1}, returning a list of each point's nearest neighbor.

\begin{CodeChunk}
\begin{CodeInput}
R> plot(waywiser::ww_build_neighbors(worldclim_testing),
+      sf::st_geometry(worldclim_testing))
R> plot(sf::st_geometry(worldclim_testing), add = TRUE)
\end{CodeInput}
\begin{figure}

{\centering \includegraphics{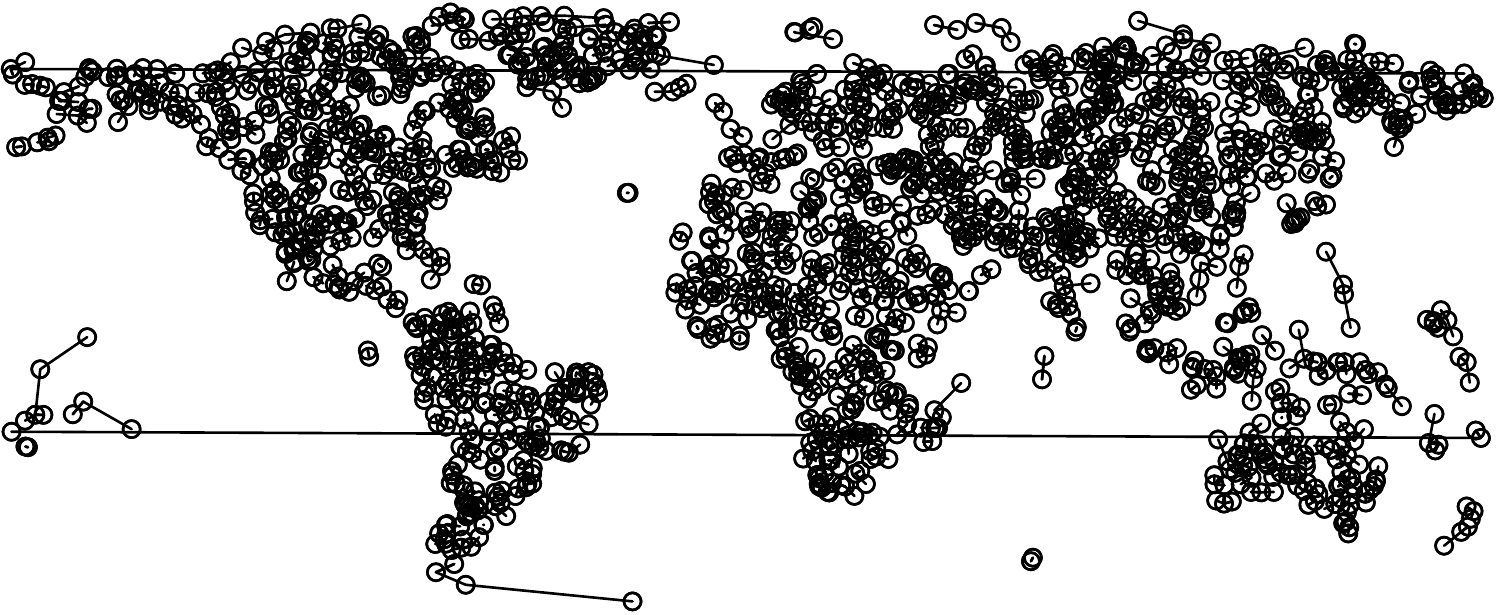} 

}

\caption[Automatically calculated spatial neighbors for points from the WorldClim simulation data]{Automatically calculated spatial neighbors for points from the WorldClim simulation data. Lines between points indicate a neighbor relationship.}\label{fig:plot-point-neighbors}
\end{figure}
\end{CodeChunk}

For calculations, this neighbor list object must be transformed into a matrix of spatial weights, \(w\). Another utility function in \pkg{waywiser}, \texttt{ww\_build\_weights()}, provides a thin wrapper around the \texttt{nb2listw()} function from \pkg{spdep}, by default producing a row-standardized spatial weights matrix.

\begin{CodeChunk}
\begin{CodeInput}
R> waywiser::ww_build_weights(guerry)
\end{CodeInput}
\begin{CodeOutput}
Characteristics of weights list object:
Neighbour list object:
Number of regions: 85 
Number of nonzero links: 420 
Percentage nonzero weights: 5.813149 
Average number of links: 4.941176 

Weights style: W 
Weights constants summary:
   n   nn S0      S1       S2
W 85 7225 85 37.2761 347.6683
\end{CodeOutput}
\end{CodeChunk}

Measures of spatial autocorrelation use this matrix in computations to estimate the relationship between variables at neighboring locations. For \pkg{waywiser}, this variable is typically assumed to be the model residual, which we will refer to as \(x\), so that for a given observation \(y_{i}\), \(x_{i} = y_{i} - \hat{y}_{i}\). By far the most popular spatial autocorrelation metric is Moran's I \citep{moran1950}, defined as:

\begin{equation}
I = \frac{n}{W}\frac{\sum{}_{i=1}^{n}\sum{}_{j=1}^{n}w_{ij}\left(x_i-\bar{x}\right)\left(x_j-\bar{x}\right)}{\sum{}_{i=1}^{n}\left(x_i-\bar{x}\right)^2}
\label{eq:global-moran}
\end{equation}

Where \(n\) is the number of observations, \(W\) the sum of all \(w_{ij}\), and \(i \neq j\). \(I\) is generally bounded \([-1, 1]\) when using row-standardized weights matrices, with positive values significantly greater than the expected value \(E(I) = -\dfrac{1}{n -1}\) indicating positive autocorrelation.

\citet{anselin1995} later expanded upon \(I\), presenting a method to estimate ``local'' \(I\) values at each observation rather than relying upon a single autocorrelation statistic to represent the entire study area:

\begin{equation}
I_{i} = \frac{\left(x_i-\bar{x}\right)\left(\sum{}_{j=1}^Nw_{ij}\left(x_j-\bar{x}\right)\right)}{m^{2}}
\label{eq:localmoran}
\end{equation}

Where \(m^2 = \dfrac{\sum{}_{i=1}^n\left(x_i-\bar{x}\right)^2}{n}\) \citep{spdep}.

A less frequently used alternative to \(I\) is Geary's \(c\) \citep{geary1954}, defined as:

\begin{equation}
c = \frac{n-1}{2W}\frac{\sum{}_{i=1}^{n}\sum{}_{j=1}^{n}w_{ij}\left(x_i-x_j\right)^{2}}{\sum{}_{i=1}^{n}\left(x_i-\bar{x}\right)^2}
\label{eq:geary-c}
\end{equation}

As with Moran's \(I\), \(c\) stated this way provides a single index of spatial autocorrelation across the entire study area. Values of \(c\) are greater than or equal to 0, with low values relative to the expected value of 1 reflecting positive autocorrelation. \citet{anselin1995} extended \(c\) in a similar manner to \(I\), providing a method to estimate local \(c\) values for each observation in a data set, with further elaboration provided in \citet{anselin2018}:

\begin{equation}
c_{i} = \sum{}_{j=1}^nw_{ij}\left(x_i-x_j\right)^{2}
\label{eq:local-c}
\end{equation}

A final metric of local spatial autocorrelation provided in \pkg{waywiser} is Getis-Ord's \(G_{i}\) \citep{getis2010, ord2010}. As with the other spatial autocorrelation metrics provided, the function implementing \(G_{i}\) in \pkg{waywiser} is a thin wrapper over a function from \pkg{spdep}, which calculates \(G_{i}\) as a standard deviate \citep{spdep, getis1996local}:

\begin{equation}
Z(G_{i})=\frac{\left[\sum_{j=1}^{n}{w_{ij}x_{j}}\right]-\left[\sum_{j=1}^{n}{w_{ij}\bar{x}}_{i}\right]}{s_i\left[\frac{\left(n-1\right)\left(\sum_{j=1}^{n}{w_{ij}^{2}}\right)-\left(\sum_{j=1}^{n}{w_{ij}}\right)^{2}}{n-1}\right]^{1/2}}, i\neq j
\label{eq:getis-ord-g}
\end{equation}

Where \(s_{i} = \sqrt{\left(\left(\sum_{j=1}^{n}{x_{j}^{2}}\right)/\left(n-1\right)\right)-\left[\bar{x}_{i}\right]^{2}},i \neq j\) and \(\bar{x}_i = \left(\sum_{j=1}^{n}{x_{j}}\right)/\left(n-1\right),i \neq j\). An extension of this metric, \(G_{i}^{*}\) removes the requirement that \(i \neq j\) by including \(i\) as a neighbor of itself, resulting in the formula \citep{spdep, getis1996local}:

\begin{equation}
Z(G_{i}^{*})=\frac{\left[\sum_{j=1}^{n}{w_{ij}x_{j}}\right]-\left[\left(\sum_{j=1}^{n}{w_{ij}}\right)\bar{x}^{*}\right]}{s^{*}\left[\frac{\left(n-1\right)\left(\sum_{j=1}^{n}{w_{ij}^{2}}\right)-\left(\sum_{j=1}^{n}{w_{ij}}\right)^{2}}{n-1}\right]^{1/2}}
\label{eq:getis-ord-gstar}
\end{equation}

Where \(s^{*} = \sqrt{\left(\left(\sum_{j=1}^{n}{x_{j}^{2}}\right)/n\right)-\bar{x}^{*2}}\) and \(\bar{x}^{*} = \left(\sum_{j=1}^{n}{x_{j}}\right)/n\). In practice, both \(G_{i}\) and \(G_{i}^{*}\) generally provide similar information \citep{getis2010}.

Much as with the model agreement metrics, spatial autocorrelation metrics are implemented in \pkg{waywiser} using the infrastructure provided by \pkg{yardstick} \citep{yardstick}, with functions prefixed with \texttt{ww\_} for autocompletion. As before, these functions take \texttt{data}, \texttt{truth}, and \texttt{estimate} as arguments, and return a tibble with \texttt{.metric}, \texttt{.estimator}, and \texttt{.estimate} for columns:

\begin{CodeChunk}
\begin{CodeInput}
R> waywiser::ww_global_moran_i(worldclim_testing,
+                             truth = response,
+                             estimate = predictions)
\end{CodeInput}
\begin{CodeOutput}
# A tibble: 1 x 3
  .metric        .estimator .estimate
  <chr>          <chr>          <dbl>
1 global_moran_i standard       0.809
\end{CodeOutput}
\end{CodeChunk}

As discussed above, by default \pkg{waywiser} will automatically create the spatial weights matrix for calculations using \texttt{ww\_build\_weights()}. To let users alter this behavior, functions for estimating spatial autocorrelation also accept an argument, \texttt{wt}, containing either the spatial weights matrix to use in calculations or a function to create the matrix from \texttt{data}:

\begin{CodeChunk}
\begin{CodeInput}
R> waywiser::ww_global_geary_c(worldclim_testing,
+                             truth = response,
+                             estimate = predictions,
+                             wt = waywiser::ww_build_weights)
\end{CodeInput}
\begin{CodeOutput}
# A tibble: 1 x 3
  .metric        .estimator .estimate
  <chr>          <chr>          <dbl>
1 global_geary_c standard       0.159
\end{CodeOutput}
\end{CodeChunk}

As the \texttt{\_vec} variants of these functions do not take an argument for \texttt{data}, \pkg{waywiser} cannot automatically create a spatial weights matrix, and one must be provided to the \texttt{wt} argument:

\begin{CodeChunk}
\begin{CodeInput}
R> waywiser::ww_global_geary_c_vec(
+   truth = worldclim_testing$response,
+   estimate = worldclim_testing$predictions,
+   wt = waywiser::ww_build_weights(worldclim_testing))
\end{CodeInput}
\begin{CodeOutput}
[1] 0.1593391
\end{CodeOutput}
\end{CodeChunk}

As previously mentioned, functions from \pkg{waywiser} are both type- and size-stable, guaranteeing that the outputs from a function will always be of a known data type and of known dimensions. For model agreement metrics and global autocorrelation statistics, this means that the output from \pkg{waywiser} will always be a tibble with one row (or, for grouped data frames, one row per group). This behavior changes for local autocorrelation metrics, however: rather than returning a single row, local autocorrelation functions return a tibble with as many rows as there are observations in \texttt{data} (or values in \texttt{truth} and \texttt{estimate}, for the \texttt{\_vec} variants). These estimates are ordered in the same order as the input data frame, meaning that the outputs from these functions can be used with \texttt{cbind()} to associate an autocorrelation estimate with its corresponding observation.

\begin{CodeChunk}
\begin{CodeInput}
R> waywiser::ww_local_moran_i(worldclim_testing,
+                            truth = response,
+                            estimate = predictions) |> head(2)
\end{CodeInput}
\begin{CodeOutput}
# A tibble: 2 x 3
  .metric       .estimator .estimate
  <chr>         <chr>          <dbl>
1 local_moran_i standard       0.290
2 local_moran_i standard       4.00 
\end{CodeOutput}
\end{CodeChunk}

Autocorrelation metrics are useful for informing qualitative assessments of model performance, with local statistics in particular being more useful for data exploration tasks than inference \citep{anselin2018}. For instance, when evaluating our linear model fit to simulation data, visually inspecting local \(I\) statistics (plotted in Figure \ref{fig:visualize-local-metrics} using \pkg{ggplot2} \citep{ggplot2} and \pkg{patchwork} \citep{patchwork}) indicate notable clustering of residuals in southern India. Local \(c\) statistics meanwhile highlight the same location as an area of concern, but also highlight several other regions in southeastern Asia, as well as areas in western Africa and South America.

\begin{CodeChunk}
\begin{figure}

{\centering \includegraphics{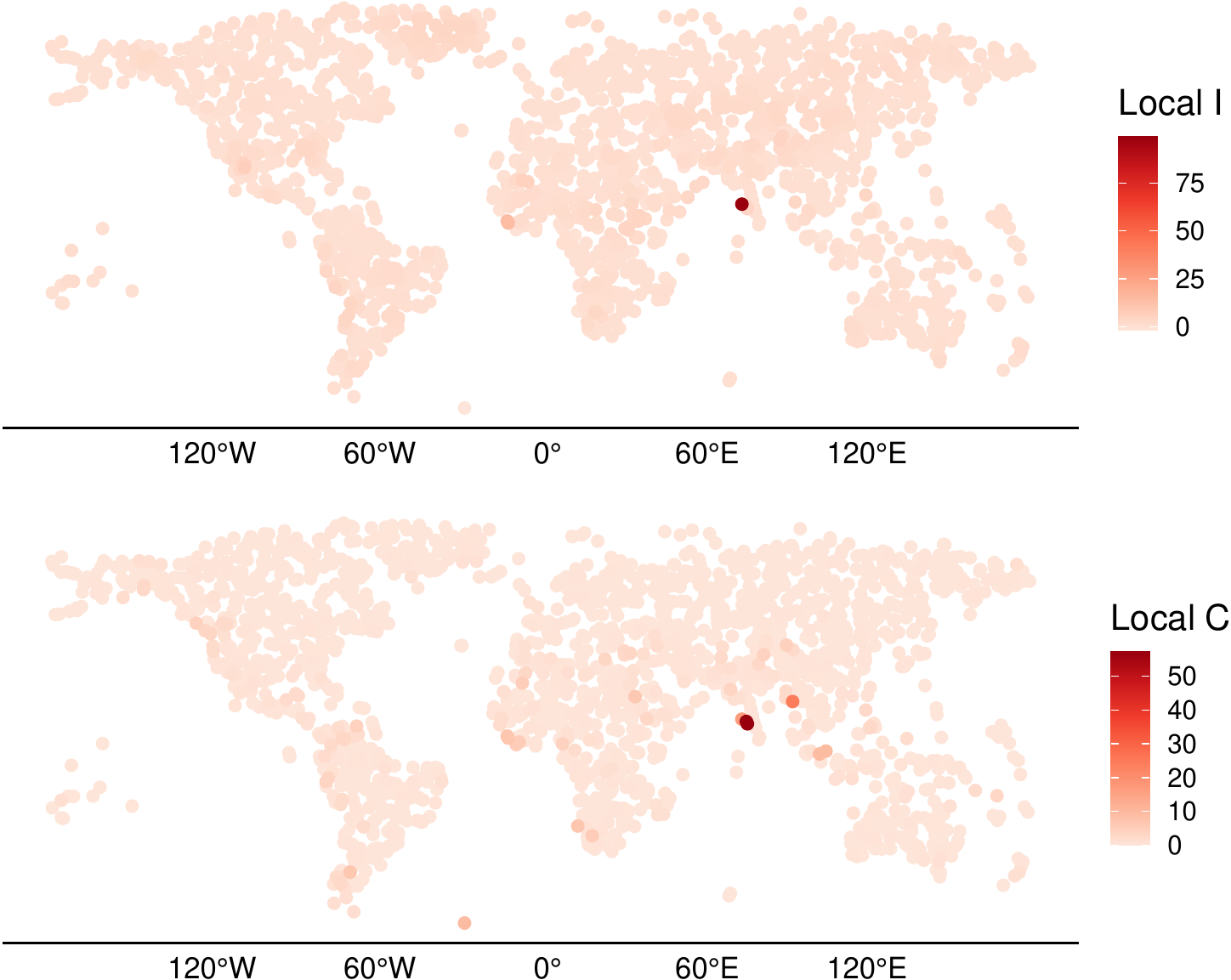} 

}

\caption[Local $I$ and $c$ values for model residuals from a linear model fit to the WorldClim simulation data]{Local $I$ and $c$ values for model residuals from a linear model fit to the WorldClim simulation data.}\label{fig:visualize-local-metrics}
\end{figure}
\end{CodeChunk}

The presence of spatial autocorrelation in model residuals suggests a misspecification in the underlying model. Similarities between regions exhibiting residual autocorrelation may help to identify missing predictors, or to explain model weaknesses.

\hypertarget{sec-multi-scale}{%
\section{Multi-scale assessment}\label{sec-multi-scale}}

Spatial models are often used to predict the most fine-grain scale of interest, with predictions then aggregated to broader scales as necessary. For instance, models of forest aboveground biomass are often used to predict biomass at individual measurement plots, and then aggregated to provide biomass estimates at regional or landscape scales \citep{johnson2022, blackard2008}. Similarly, models of nematode biomass are used to estimate biomass at fine-grained resolutions, with predictions then aggregated to produce global estimates \citep{vandenhoogen2019}. Unfortunately, model performance estimates cannot be assumed to be consistent across spatial scales \citep{nelson2009}. As a result, separate performance metrics must be calculated for each relevant scale of interest.

To this end, \citet{riemann2010} introduced an assessment protocol for evaluating model performance across multiple scales of aggregation. This method requires calculating multiple performance estimates from model predictions and measured values aggregated to multiple grids of regular polygons, in order to assess how model performance varies spatially and across multiple scales. This approach is implemented in \pkg{waywiser} as the function \texttt{ww\_multi\_scale()}. As with the model assessment functions already described, this function accepts the arguments \texttt{data}, \texttt{truth}, and \texttt{estimate}, as well as a new argument \texttt{metrics} which accepts lists of metric functions (or outputs from \texttt{yardstick::metric\_set()}) to calculate at each scale of aggregation. Additional arguments can be passed via \texttt{...} to \texttt{sf::st\_make\_grid()} in order to assess predictions aggregated to a grid of evenly spaced regular polygons. For example, the following code evaluates the RMSE and Willmott's \(d_1\) values for the WorldClim model aggregated using a 2-by-2, 5-by-5, and 10-by-10 grid:

\begin{CodeChunk}
\begin{CodeInput}
R> (multi_scale_output <- waywiser::ww_multi_scale(
+   worldclim_testing,
+   truth = response,
+   estimate = predictions,
+   metrics = list(yardstick::rmse, waywiser::ww_willmott_d1), 
+   n = c(2, 5, 10)))
\end{CodeInput}
\begin{CodeOutput}
# A tibble: 6 x 6
  .metric     .estimator .estimate .grid_args       .grid          .notes  
  <chr>       <chr>          <dbl> <list>           <list>         <list>  
1 rmse        standard      0.0299 <tibble [1 x 1]> <sf [4 x 5]>   <tibble>
2 willmott_d1 standard      0.586  <tibble [1 x 1]> <sf [4 x 5]>   <tibble>
3 rmse        standard      0.0650 <tibble [1 x 1]> <sf [25 x 5]>  <tibble>
4 willmott_d1 standard      0.759  <tibble [1 x 1]> <sf [25 x 5]>  <tibble>
5 rmse        standard      0.0761 <tibble [1 x 1]> <sf [100 x 5]> <tibble>
6 willmott_d1 standard      0.810  <tibble [1 x 1]> <sf [100 x 5]> <tibble>
\end{CodeOutput}
\end{CodeChunk}

Note that this function requires \texttt{data} be an \texttt{sf} object with point geometries from the \pkg{sf} package, in order to ensure that the data's coordinate reference system and units are properly handled when assigning observations to each grid cell.

The output from \texttt{ww\_multi\_scale()} will typically have one row for each combination of metric and aggregation level. As with functions for model agreement metrics, however, if \texttt{data} is a grouped data frame produced by \texttt{group\_by()}, this function will instead calculate metrics for each group independently, resulting in an output with one row per combination of unique grouping, metric, and aggregation:

\begin{CodeChunk}
\begin{CodeInput}
R> waywiser::ww_multi_scale(
+   dplyr::group_by(worldclim_testing, group),
+   truth = response,
+   estimate = predictions,
+   metrics = list(waywiser::ww_willmott_d1), 
+   n = 2)
\end{CodeInput}
\begin{CodeOutput}
# A tibble: 2 x 7
  group .metric     .estimator .estimate .grid_args       .grid        .notes  
  <int> <chr>       <chr>          <dbl> <list>           <list>       <list>  
1     1 willmott_d1 standard       0.776 <tibble [1 x 1]> <sf [8 x 6]> <tibble>
2     2 willmott_d1 standard       0.408 <tibble [1 x 1]> <sf [8 x 6]> <tibble>
\end{CodeOutput}
\end{CodeChunk}

In addition to the familiar \texttt{.metric}, \texttt{.estimator}, and \texttt{.estimate} columns, \texttt{ww\_multi\_scale()} returns three new columns. The \texttt{.grid\_args} column is a list of tibbles from the \pkg{tibble} package \citep{tibble}, containing the arguments used to construct the grid via \texttt{sf::st\_make\_grid()}. This column can be unpacked, for example via the \texttt{unnest()} function in \pkg{tidyr} \citep{tidyr}, to add these arguments as columns to the output table:

\begin{CodeChunk}
\begin{CodeInput}
R> tidyr::unnest(multi_scale_output, .grid_args)
\end{CodeInput}
\begin{CodeOutput}
# A tibble: 6 x 6
  .metric     .estimator .estimate     n .grid          .notes          
  <chr>       <chr>          <dbl> <dbl> <list>         <list>          
1 rmse        standard      0.0299     2 <sf [4 x 5]>   <tibble [0 x 2]>
2 willmott_d1 standard      0.586      2 <sf [4 x 5]>   <tibble [0 x 2]>
3 rmse        standard      0.0650     5 <sf [25 x 5]>  <tibble [0 x 2]>
4 willmott_d1 standard      0.759      5 <sf [25 x 5]>  <tibble [0 x 2]>
5 rmse        standard      0.0761    10 <sf [100 x 5]> <tibble [0 x 2]>
6 willmott_d1 standard      0.810     10 <sf [100 x 5]> <tibble [0 x 2]>
\end{CodeOutput}
\end{CodeChunk}

This is particularly convenient when visualizing the outputs from this process:

\begin{CodeChunk}
\begin{CodeInput}
R> tidyr::unnest(multi_scale_output, .grid_args) |> 
+   ggplot2::ggplot(ggplot2::aes(n^2, .estimate, color = .metric)) + 
+   ggplot2::geom_line() +
+   ggplot2::scale_x_continuous(
+     name = "Number of grid cells",
+     breaks = (c(2, 5, 10)^2)) + 
+   ggplot2::facet_wrap(~ .metric)
\end{CodeInput}
\begin{figure}

{\centering \includegraphics{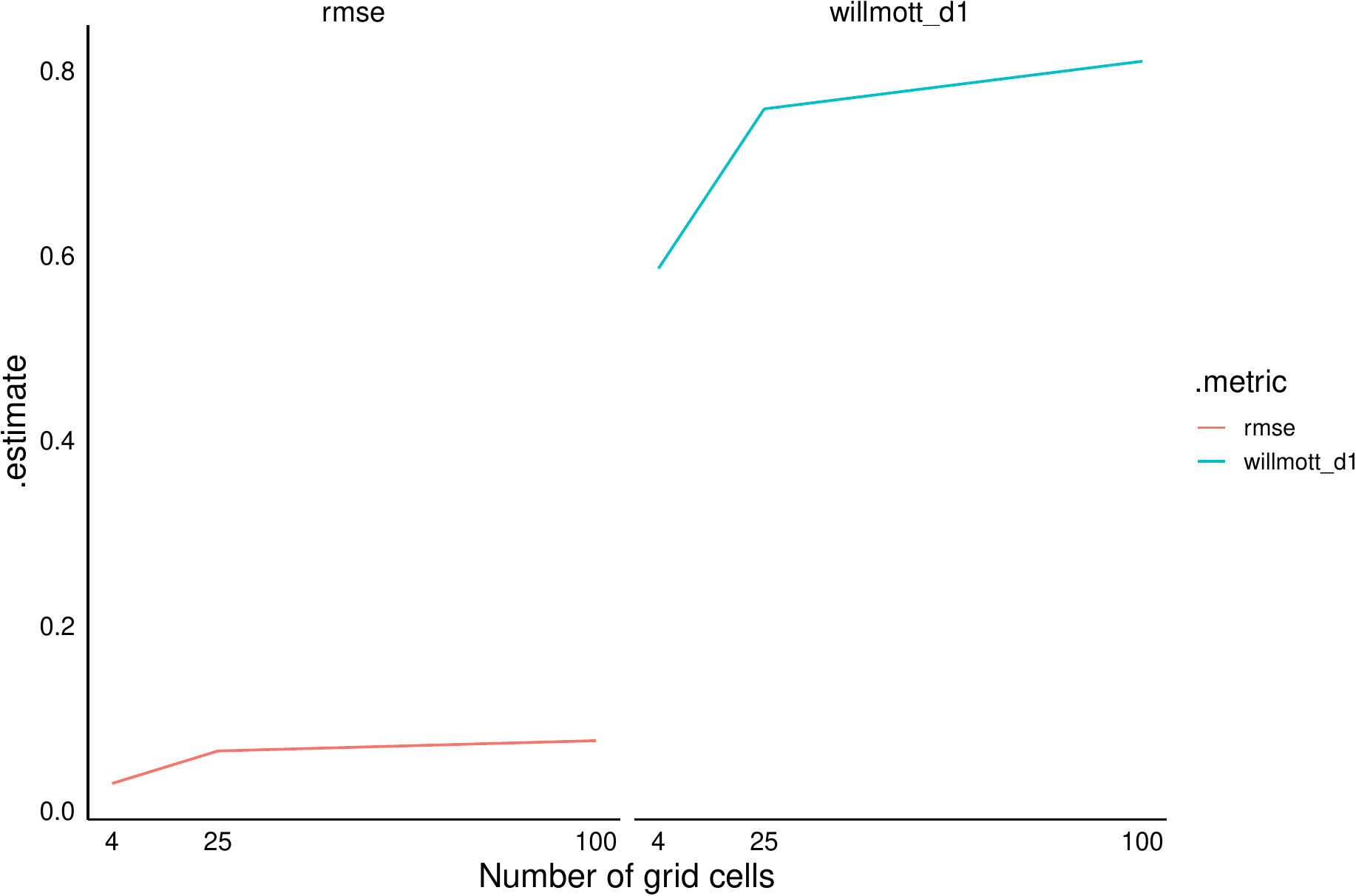} 

}

\caption[RMSE and Willmott $d_{1}$ estimates for a linear model fit to the WorldClim simulation data, aggregated to multiple spatial scales]{RMSE and Willmott $d_{1}$ estimates for a linear model fit to the WorldClim simulation data, aggregated to multiple spatial scales. The "number of grid cells" refers to the number of partitions data was aggregated into; more grid cells indicate a finer-grained level of aggregation.}\label{fig:vis-grid-args}
\end{figure}
\end{CodeChunk}

The \texttt{.grid} column meanwhile is a list of \texttt{sf} objects, each containing the actual polygons used to aggregate predictions and observed values before calculating metrics. These objects also contain the aggregated values of \texttt{.truth} and \texttt{.estimate}, as well as a count of the number of non-missing observations for each of \texttt{.truth} and \texttt{.estimate} contained in the polygon.

\begin{CodeChunk}
\begin{CodeInput}
R> multi_scale_output$.grid[[2]]
\end{CodeInput}
\begin{CodeOutput}
Simple feature collection with 4 features and 4 fields
Geometry type: POLYGON
Dimension:     XY
Bounding box:  xmin: -178.6972 ymin: -59.39802 xmax: 179.6074 ymax: 83.85131
Geodetic CRS:  +proj=longlat +datum=WGS84 +no_defs
     .truth .truth_count .estimate .estimate_count
1 0.3221379          790 0.3613896             790
2 0.4078587         1191 0.3860225            1191
3 0.3148534           10 0.3539468              10
4 0.4041676            9 0.4096796               9
                        geometry
1 POLYGON ((-178.6972 -59.398...
2 POLYGON ((0.4550702 -59.398...
3 POLYGON ((-178.6972 12.2266...
4 POLYGON ((0.4550702 12.2266...
\end{CodeOutput}
\end{CodeChunk}

This data is often useful to visualize the spatial distribution of error:

\begin{CodeChunk}
\begin{CodeInput}
R> multi_scale_output$.grid[[6]] |> 
+   tidyr::drop_na() |> 
+   ggplot2::ggplot(ggplot2::aes(fill = .truth - .estimate)) + 
+   ggplot2::geom_sf() + 
+   ggplot2::scale_fill_distiller(palette = "YlOrRd", direction = 1)
\end{CodeInput}
\begin{figure}

{\centering \includegraphics{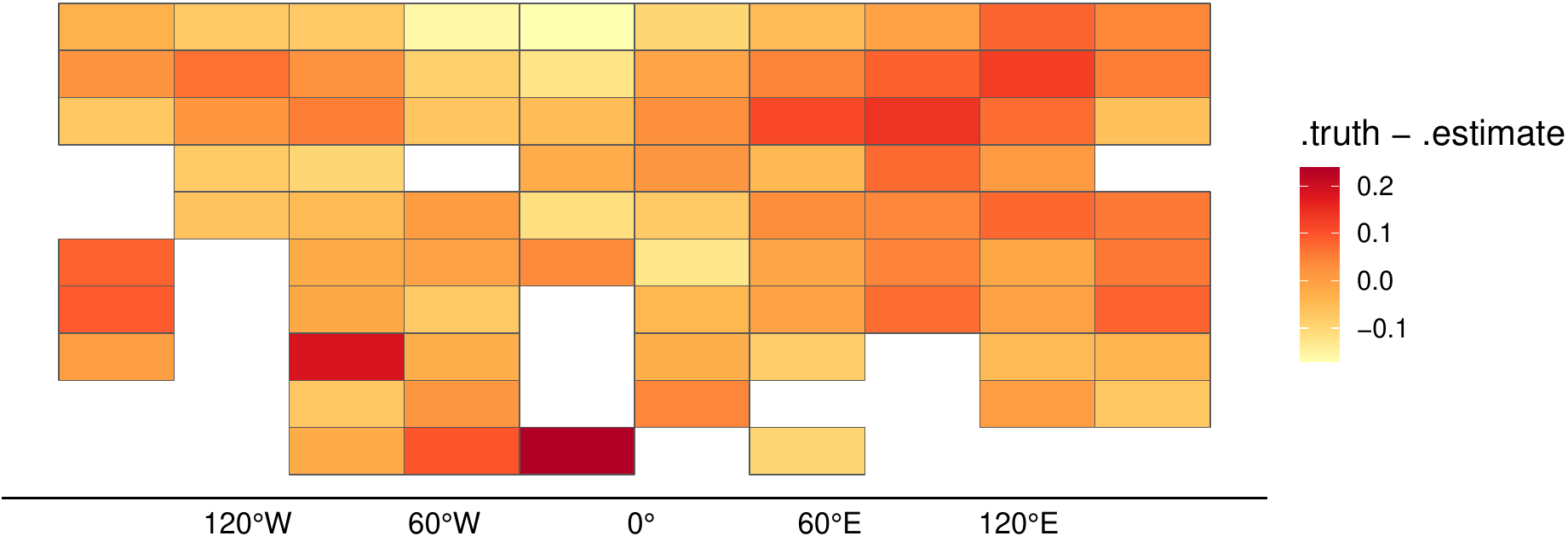} 

}

\caption[Model errors after aggregating predictions and observed values to a 10-by-10 grid]{Model errors after aggregating predictions and observed values to a 10-by-10 grid.}\label{fig:vis-grid}
\end{figure}
\end{CodeChunk}

The final \texttt{.notes} column contains diagnostic information, including information on which (if any) observations fell outside the boundaries of the grid.

When not all use cases for a model are known, this approach of systematically aggregating predictions and observed values to grids of regular polygons can help provide useful performance estimates for future users who will aggregate predictions to their boundaries of interest. If future use cases are well understood, however, it is often more useful to assess model performance when aggregating to those boundaries directly; for instance, it often makes sense to evaluate model performance when aggregating predictions to administrative boundaries such as along town or regional polygons, in addition to other assessments.

For this reason, \texttt{ww\_multi\_scale()} allows users to provide their own pre-computed polygons to the \texttt{grids} argument. Predictions and observed values will then be aggregated to the provided polygons. Other than this, the function interface and returned values are identical:

\begin{CodeChunk}
\begin{CodeInput}
R> waywiser::ww_multi_scale(
+   worldclim_testing,
+   truth = response,
+   estimate = predictions,
+   metrics = list(yardstick::rmse, waywiser::ww_willmott_d1),
+   grids = list(sf::st_make_grid(worldclim_testing)))
\end{CodeInput}
\begin{CodeOutput}
# A tibble: 2 x 6
  .metric     .estimator .estimate .grid_args       .grid          .notes  
  <chr>       <chr>          <dbl> <list>           <list>         <list>  
1 rmse        standard      0.0761 <tibble [0 x 0]> <sf [100 x 5]> <tibble>
2 willmott_d1 standard      0.810  <tibble [0 x 0]> <sf [100 x 5]> <tibble>
\end{CodeOutput}
\end{CodeChunk}

Note that when using pre-computed polygons \texttt{.grid\_args} is a tibble with 0 rows, as no arguments were passed to create the grid.

A challenge with requiring \texttt{sf} objects for input data is that it requires all observations be loaded into memory before any aggregations can be calculated, which poses problems for modeling efforts with large numbers of relatively small observation units such as high-resolution landscape or global models. Such efforts typically rely on raster formats for efficiency, and avoid transforming rasters into vector representations whenever possible. As such, \texttt{ww\_multi\_scale()} can also be used to calculate performance metrics using raster inputs. If \texttt{data} is \texttt{NULL} or missing, users can provide \texttt{SpatRaster} objects from the \pkg{terra} package \citep{terra} to both \texttt{truth} and \texttt{estimate} in order to perform a multi-scale assessment. Aggregation for this method is performed using the \pkg{exactextractr} package \citep{exactextractr} for memory and computational efficiency. Other than the slightly different interface and using \pkg{exactextractr} for aggregation, this method of \texttt{ww\_multi\_scale()} works identically and returns an identical output to the \texttt{sf} method.

\begin{CodeChunk}
\begin{CodeInput}
R> r1 <- matrix(nrow = 10, ncol = 10)
R> r1[] <- 1
R> r1 <- terra::rast(r1)
R> r2 <- matrix(nrow = 10, ncol = 10)
R> r2[] <- 2
R> r2 <- terra::rast(r2)
R> waywiser::ww_multi_scale(truth = r1, estimate = r2, n = 1)
\end{CodeInput}
\begin{CodeOutput}
# A tibble: 2 x 6
  .metric .estimator .estimate .grid_args       .grid        .notes          
  <chr>   <chr>          <dbl> <list>           <list>       <list>          
1 rmse    standard           1 <tibble [1 x 1]> <sf [1 x 5]> <tibble [0 x 2]>
2 mae     standard           1 <tibble [1 x 1]> <sf [1 x 5]> <tibble [0 x 2]>
\end{CodeOutput}
\end{CodeChunk}

\hypertarget{sec-aoa}{%
\section{Area of applicability}\label{sec-aoa}}

A final set of functions in \pkg{waywiser} aim to help users assess if their model can be trusted to generalize to new observations. By calculating how similar new observations are in predictor space to the data used to train a model, it is possible to determine if new observations are within the ``applicability domain'' of a model and are likely to be well-represented by the model's predictions \citep{netzeva2005}. While these methods are not explicitly spatial, the presence of spatial autocorrelation in model predictors makes it more likely that using a model to extrapolate geographically will also require the model to extrapolate in predictor space, producing worse predictions as the extrapolation distance increases. As such, these techniques are particularly useful when predicting into regions with little data for model training and assessment.

\citet{meyer2021} introduce a new applicability domain methodology built upon a ``dissimilarity index'', \(DI\), representing the Euclidean distance in predictor space between a point and its nearest neighbor scaled by the average such distance in the data used to train a model. Using variables which have been scaled and centered, then weighted by variable importance scores, the \(DI\) for an observation \(k\) is calculated as:

\begin{equation}
\begin{aligned}
d\left(a, b\right) &= \sqrt{\sum{}_{i=1}^{p}\left(a_i - b_i\right)^2}\\
d_k &= \operatorname*{arg\,min}_{z}d(k, z)\\
DI_k &= \dfrac{d_k}{\bar{d}}
\end{aligned}
\label{eq:di}
\end{equation}

Where \(p\) is the number of predictors used in fitting a model, \(a_i\) and \(b_i\) the scaled and weighted predictor values for observations \(a\) and \(b\), \(z\) observations in the data used to train the model, and \(\bar{d}\) the mean \(d\) for all pairs of observations in the data used to train the model.

These \(DI\) values are useful in their own right to characterize the similarity of new observations to those used to train a model, and increasing \(DI\) often correlates with increasing prediction error \citep{meyer2021}. \citet{meyer2021} also propose a thresholding method to calculate a boolean ``area of applicability'' (AOA), defining points with a \(DI_k\) greater than the 75th percentile \(DI\) value plus 1.5 times the IQR of \(DI\) values in training data as beyond the model's AOA, and therefore likely to have greater prediction error than reported for test set observations.

Functions to calculate \(DI\) and AOA were first implemented in \pkg{CAST} \citep{CAST}, with a focus on supporting models fit using the \pkg{caret} modeling framework \citep{caret}. In \pkg{waywiser}, the \texttt{ww\_area\_of\_applicability()} function provides a framework-agnostic interface for calculating \(DI\) and AOA, with additional support for workflows using the tidymodels framework (Section \ref{sec-interop}) \citep{tidymodels}. The interface of \texttt{ww\_area\_of\_applicability()} is inspired by the \pkg{applicable} package \citep{applicable}, and mimics common model-fitting functions such as \texttt{lm()}. A standard call involves providing the model formula used, training and testing data sets, and variable importance scores:

\begin{CodeChunk}
\begin{CodeInput}
R> (aoa <- waywiser::ww_area_of_applicability(
+   formula(worldclim_model),
+   worldclim_training,
+   testing = worldclim_testing,
+   importance = vip::vi_model(worldclim_model)))
\end{CodeInput}
\begin{CodeOutput}
# Predictors:
   4
Area-of-applicability threshold:
   0.103787
\end{CodeOutput}
\end{CodeChunk}

An equivalent call removes the formula argument, and instead passes the training and testing data, subset to include only the variables used to fit the model:

\begin{CodeChunk}
\begin{CodeInput}
R> waywiser::ww_area_of_applicability(
+   as.data.frame(worldclim_training)[1:4],
+   testing = as.data.frame(worldclim_testing)[1:4],
+   importance = vip::vi_model(worldclim_model))
\end{CodeInput}
\begin{CodeOutput}
# Predictors:
   4
Area-of-applicability threshold:
   0.103787
\end{CodeOutput}
\end{CodeChunk}

As demonstrated, \texttt{ww\_area\_of\_applicability()} natively accepts variable importance scores returned by the \pkg{vip} package \citep{vip}. The \texttt{importance} argument will also accept any data frame with columns named \texttt{term} and \texttt{estimate}, containing (respectively) the variable name and importance estimate, allowing the use of any method for calculating variable importance scores.

Unlike other functions in \pkg{waywiser}, the output from \texttt{ww\_area\_of\_applicability()} is a custom class ``\texttt{ww\_area\_of\_applicability}'' which inherits from the ``\texttt{hardhat\_model}'' and ``\texttt{hardhat\_scalar}'' classes from the \pkg{hardhat} package. This class can be thought of as being a model object, like that returned by \texttt{lm()}, and as such implements a \texttt{predict()} method for calculating \(DI\) and AOA for new observations. This function returns a tibble with two columns: \texttt{di}, containing \(DI\) values for each new observation, and \texttt{aoa}, a boolean indicating if the observation is within (\texttt{TRUE}) or outside of (\texttt{FALSE}) the AOA:

\begin{CodeChunk}
\begin{CodeInput}
R> predict(aoa, worldclim_testing)
\end{CodeInput}
\begin{CodeOutput}
# A tibble: 2,000 x 2
       di aoa  
    <dbl> <lgl>
 1 0.0676 TRUE 
 2 0.109  FALSE
 3 0.0264 TRUE 
 4 0.0382 TRUE 
 5 0.0398 TRUE 
 6 0.0243 TRUE 
 7 0      TRUE 
 8 0.0462 TRUE 
 9 0.0285 TRUE 
10 0.0178 TRUE 
# ... with 1,990 more rows
\end{CodeOutput}
\end{CodeChunk}

Observations with missing values will have an \texttt{NA} for both \texttt{di} and \texttt{aoa}, guaranteeing that the number of rows returned will always match the number of rows in the \texttt{newdata} object. As such, these predictions can be safely combined with predictor values via \texttt{cbind()} and similar functions, making visualization and analysis easier.

\begin{CodeChunk}
\begin{CodeInput}
R> library("patchwork")
R> (cbind(worldclim_testing, predict(aoa, worldclim_testing)) |> 
+     ggplot2::ggplot(ggplot2::aes(color = di)) + 
+     ggplot2::geom_sf(alpha = 0.7) + 
+     ggplot2::scale_color_distiller(palette = "Reds", direction = 1)) /
+   (cbind(worldclim_testing, predict(aoa, worldclim_testing)) |> 
+      ggplot2::ggplot(ggplot2::aes(color = aoa)) + 
+      ggplot2::geom_sf(alpha = 0.7))
\end{CodeInput}
\begin{figure}

{\centering \includegraphics{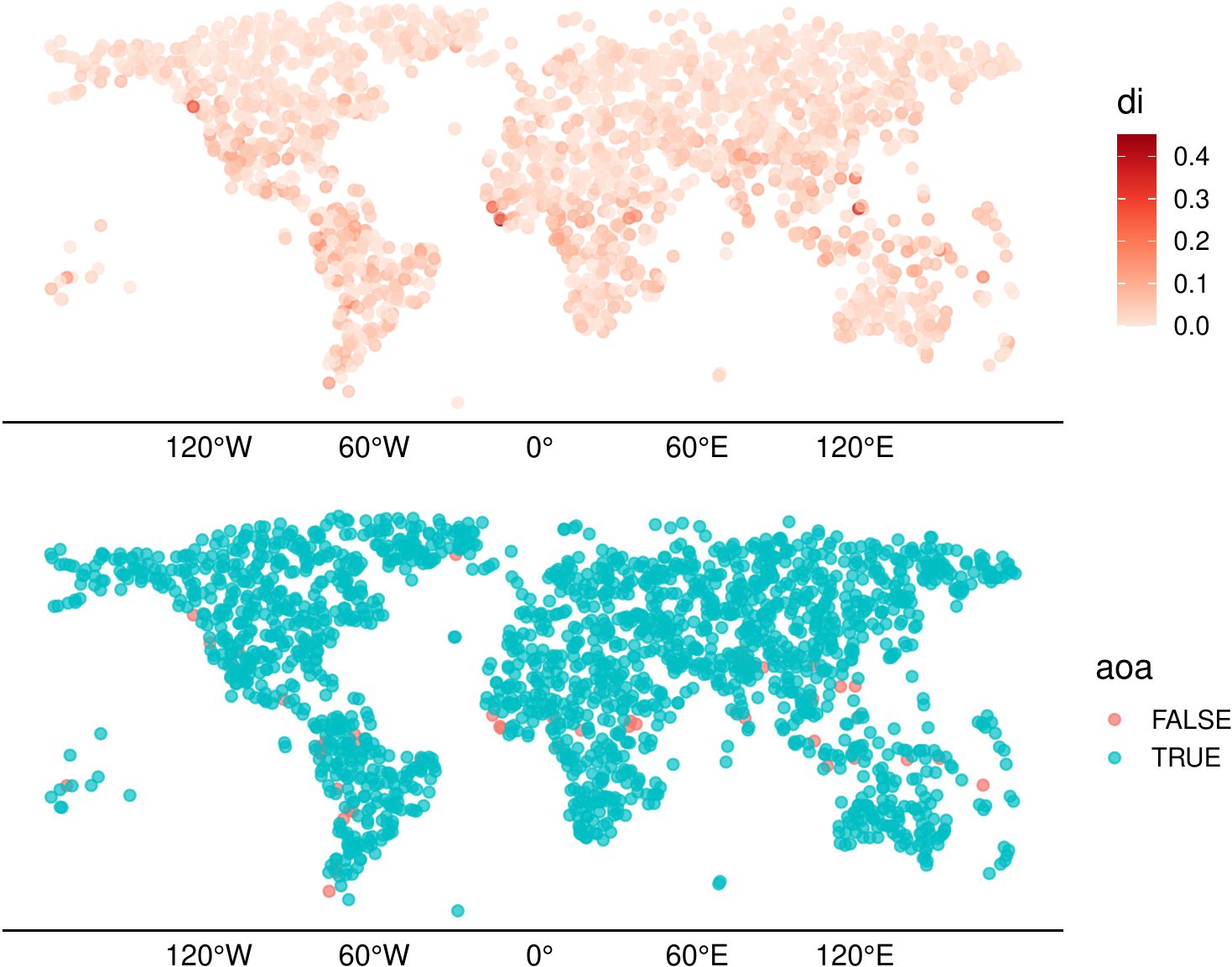} 

}

\caption[$DI$ and AOA for for a linear model fit to the WorldClim simulation data]{$DI$ and AOA for for a linear model fit to the WorldClim simulation data. Areas with a higher $DI$ are poorly represented in the data used to train the model, and are likely to fall outside the model's AOA (`FALSE` in the lower graph).}\label{fig:aoa-graph}
\end{figure}
\end{CodeChunk}

Finally, \texttt{ww\_area\_of\_applicability()} also supports calculating the AOA using data splits from cross-validation, as opposed to distinct training and testing sets. Any \texttt{rset} object, such as those produced by the \pkg{rsample} or \pkg{spatialsample} packages \citep{rsample, spatialsample}, can be used to calculate the AOA. This method calculates \(d_k\) as the distance between a point in the assessment set and its nearest neighbor in the analysis set. A full demonstration is included in Section \ref{sec-interop}.

When using cross-validation splits in the place of test set data, \pkg{waywiser} differs slightly from the implementation in \pkg{CAST}. Whereas \pkg{CAST} performs scaling and centering using all observations, and calculates \(\bar{d}\) as the mean distance between all observations across all folds, \pkg{waywiser} rescales the analysis and assessment sets for each fold of cross validation separately, and calculates a per-fold \(\bar{d}\) as the mean distance between observations in only the analysis set. These changes aim to limit data leakage between the analysis and assessment data sets, ensuring that assessment data is not used to determine parameters for centering and scaling analysis data or to calculate \(\bar{d}\). In practice, this means that \pkg{waywiser} calculates a somewhat higher \(d_k\) than \pkg{CAST}, which results in a slightly higher \(DI_k\) and threshold value. For predictions, \pkg{waywiser} scales and centers the entire data set as a whole, and sets \(\bar{d}\) to the mean \(\bar{d}\) across folds, under the assumption that the final model in use has been retrained using the entire data set.

\hypertarget{sec-interop}{%
\section{Interoperability}\label{sec-interop}}

As previous sections have demonstrated, \pkg{waywiser} is designed to be useful for most spatial modeling workflows, no matter what models or frameworks are used. Functions in \pkg{waywiser} are size- and type-stable, returning outputs of the same data type and predictable dimensions regardless of the input arguments used. Functions rely upon well-established data types for inputs and outputs, using standard data frames and vectors where appropriate and relying on the popular \pkg{sf} package for spatial data classes. By relying upon standard classes and returning predictable outputs, \pkg{waywiser} aims to integrate easily with as many modeling workflows as possible.

However, \pkg{waywiser} is particularly well suited for workflows leveraging the tidymodels ecosystem of modeling packages \citep{tidymodels}. These packages provide a consistent, user-friendly interface for common modeling tasks, with the aim of making it easy for users to follow statistical best practices simply by using package functions in the most straightforward way. Functions in \pkg{waywiser} are designed to be automatically compatible with tidymodels packages. To demonstrate this, we will first split our data into folds for cross-validation, using a 10-fold spatial clustering cross-validation approach, as implemented in \pkg{spatialsample} \citep{spatialsample}:

\begin{CodeChunk}
\begin{CodeInput}
R> library("tidymodels") |> 
+   suppressPackageStartupMessages()
R> (worldclim_resamples <- 
+     spatialsample::spatial_clustering_cv(worldclim_training))
\end{CodeInput}
\begin{CodeOutput}
#  10-fold spatial cross-validation 
# A tibble: 10 x 2
   splits              id    
   <list>              <chr> 
 1 <split [7180/820]>  Fold01
 2 <split [7338/662]>  Fold02
 3 <split [6757/1243]> Fold03
 4 <split [7254/746]>  Fold04
 5 <split [6900/1100]> Fold05
 6 <split [7226/774]>  Fold06
 7 <split [7288/712]>  Fold07
 8 <split [7452/548]>  Fold08
 9 <split [7261/739]>  Fold09
10 <split [7344/656]>  Fold10
\end{CodeOutput}
\end{CodeChunk}

Each row of \texttt{worldclim\_resamples} contains a single cross-validation iteration, with data split into analysis and assessment sets based on spatial location; each observation is assigned to precisely one assessment set. We can use functions from \pkg{workflows}, \pkg{parsnip}, \pkg{tune}, and \pkg{yardstick} from the tidymodels ecosystem to then fit separate linear models to each of these analysis sets, and evaluate them against their respective assessment set \citep{workflows, parsnip, tune, yardstick}. Note that we can easily integrate model assessment functions from \pkg{waywiser} into this workflow, as a result of these functions extending infrastructure from \pkg{yardstick}:

\begin{CodeChunk}
\begin{CodeInput}
R> workflow(response ~ bio2 + bio10 + bio13 + bio19) |> 
+   add_model(linear_reg()) |> 
+   fit_resamples(worldclim_resamples,
+                 metrics = metric_set(rmse, mae, waywiser::ww_willmott_d1)) |> 
+   collect_metrics()
\end{CodeInput}
\begin{CodeOutput}
# A tibble: 3 x 6
  .metric     .estimator   mean     n std_err .config             
  <chr>       <chr>       <dbl> <int>   <dbl> <chr>               
1 mae         standard   0.0968    10 0.00875 Preprocessor1_Model1
2 rmse        standard   0.120     10 0.0106  Preprocessor1_Model1
3 willmott_d1 standard   0.642     10 0.0465  Preprocessor1_Model1
\end{CodeOutput}
\end{CodeChunk}

We can similarly use model assessment functions from \pkg{waywiser} for other purposes, for instance to automatically evaluate hyperparameters for a random forest model using the \pkg{dials} package \citep{dials}:

\begin{CodeChunk}
\begin{CodeInput}
R> rf_workflow <- workflow(response ~ bio2 + bio10 + bio13 + bio19) |> 
+   add_model(rand_forest("regression", mtry = tune(), 
+                         trees = tune(), min_n = tune()))
R> rf_parameters <- extract_parameter_set_dials(rf_workflow) |> 
+   finalize(worldclim_resamples)
R> rf_res <- rf_workflow |> 
+   tune_grid(grid = grid_latin_hypercube(rf_parameters, size = 9),
+             resamples = worldclim_resamples,
+             metrics = metric_set(waywiser::ww_willmott_d1))
R> collect_metrics(rf_res)
\end{CodeInput}
\begin{CodeOutput}
# A tibble: 9 x 9
   mtry trees min_n .metric     .estimator  mean     n std_err .config          
  <int> <int> <int> <chr>       <chr>      <dbl> <int>   <dbl> <chr>            
1     2   971    32 willmott_d1 standard   0.935    10 0.0102  Preprocessor1_Mo~
2     1   325    17 willmott_d1 standard   0.916    10 0.0129  Preprocessor1_Mo~
3     2   482     7 willmott_d1 standard   0.945    10 0.00945 Preprocessor1_Mo~
4     2    95    26 willmott_d1 standard   0.936    10 0.0101  Preprocessor1_Mo~
5     1  1657    29 willmott_d1 standard   0.910    10 0.0144  Preprocessor1_Mo~
6     1  1522     5 willmott_d1 standard   0.926    10 0.0123  Preprocessor1_Mo~
7     2   728    23 willmott_d1 standard   0.938    10 0.00985 Preprocessor1_Mo~
8     1  1779    36 willmott_d1 standard   0.907    10 0.0143  Preprocessor1_Mo~
9     2  1134    14 willmott_d1 standard   0.942    10 0.00954 Preprocessor1_Mo~
\end{CodeOutput}
\begin{CodeInput}
R> select_best(rf_res)
\end{CodeInput}
\begin{CodeOutput}
# A tibble: 1 x 4
   mtry trees min_n .config             
  <int> <int> <int> <chr>               
1     2   482     7 Preprocessor1_Model3
\end{CodeOutput}
\end{CodeChunk}

Having found the optimal hyperparameters for the random forest, we can then fit our tuned model to our full set of training data and use functions from \pkg{yardstick} and \pkg{waywiser} to evaluate predictions against the test set:

\begin{CodeChunk}
\begin{CodeInput}
R> tuned_rf_workflow <- rf_workflow |> 
+   finalize_workflow(select_best(rf_res)) |> 
+   fit(worldclim_training)
R> worldclim_testing$predictions <- predict(tuned_rf_workflow,
+                                          worldclim_testing)$.pred
R> metrics <- metric_set(waywiser::ww_willmott_d1,
+                       waywiser::ww_agreement_coefficient)
R> metrics(worldclim_testing, response, predictions)
\end{CodeInput}
\begin{CodeOutput}
# A tibble: 2 x 3
  .metric               .estimator .estimate
  <chr>                 <chr>          <dbl>
1 willmott_d1           standard       0.983
2 agreement_coefficient standard       0.998
\end{CodeOutput}
\end{CodeChunk}

Finally, as alluded to in Section \ref{sec-aoa}, we may also use \texttt{ww\_area\_of\_applicability()} with our cross-validation object to estimate the AOA of this random forest. In order to do so, we will first estimate our variable importance scores through \texttt{vip::vi\_permute()}:

\begin{CodeChunk}
\begin{CodeInput}
R> d1_wrapper <- function(actual, predicted) {
+   waywiser::ww_willmott_d1_vec(actual, predicted)
+ }
R> pred_wrapper <- function(object, newdata) {
+   object |> 
+     predict(newdata) |> 
+     ranger::predictions()
+ }
R> importance <- vip::vi_permute(
+   extract_fit_engine(tuned_rf_workflow),
+   train = as.data.frame(worldclim_training)[c(1:4, 6)],
+   target = "response",
+   metric = d1_wrapper,
+   smaller_is_better = FALSE,
+   pred_wrapper = pred_wrapper)
\end{CodeInput}
\end{CodeChunk}

We are then able to calculate our area of applicability by passing our cross-validation object, model formula, and importance scores to \texttt{ww\_area\_of\_applicability()}:

\begin{CodeChunk}
\begin{CodeInput}
R> waywiser::ww_area_of_applicability(
+   worldclim_resamples,
+   response ~ bio2 + bio10 + bio13 + bio19,
+   importance = importance)
\end{CodeInput}
\begin{CodeOutput}
# Predictors:
   4
Area-of-applicability threshold:
   0.1791143
\end{CodeOutput}
\end{CodeChunk}

In this way, \pkg{waywiser} functions are designed to integrate naturally with packages in the tidymodels ecosystem, extending the consistent user-friendly interfaces of those packages for spatial model assessment tasks. However, as emphasized at the start of this section, \pkg{waywiser} functions are designed to be framework-agnostic and also accept standard data frames and vectors as inputs, while returning standard data frames and vectors as outputs wherever possible. In this way, \pkg{waywiser} aims to be maximally interoperable with as many modeling workflows as possible.

\hypertarget{conclusion}{%
\section{Conclusion}\label{conclusion}}

The \pkg{waywiser} package provides an ergonomic toolkit for assessing spatial models, with a focus on providing a consistent interface to multiple well-established assessment methods. Functions provided by \pkg{waywiser} include a number of model assessment metrics, an approach for assessing model performance when aggregating predictions across multiple spatial scales, and an approach for calculating the applicability domain of a model. These functions accept and, where possible, return values as standard data frames and vectors, making them compatible with a wide swath of modeling workflows. Additional features make it particularly easy to use \pkg{waywiser} in combination with packages from the tidymodels ecosystem. Future directions for the package will include the addition of new assessment metrics and computational speedups. Release versions of \pkg{waywiser} are available from the Comprehensive \proglang{R} Archive Network (CRAN) at \url{https://CRAN.R-project.org/package=waywiser}, while development versions may be downloaded from GitHub at \url{https://github.com/ropensci/waywiser}.

\hypertarget{acknowledgments}{%
\section*{Acknowledgments}\label{acknowledgments}}
\addcontentsline{toc}{section}{Acknowledgments}

Initial development of \pkg{waywiser} was supported by Posit, PBC. Version 0.3.0 of \pkg{waywiser} was reviewed for rOpenSci by Dr.~Virgilio Gómez-Rubio and Dr.~Jakub Nowosad, whose feedback greatly improved the package. Lucas Johnson provided valuable feedback on the area of applicability implementation and the multi-scale assessment workflow.

\renewcommand\refname{References}
\bibliography{paper.bib}

\begin{thebibliography}{54}
\newcommand{\enquote}[1]{``#1''}
\providecommand{\natexlab}[1]{#1}
\providecommand{\url}[1]{\texttt{#1}}
\providecommand{\urlprefix}{URL }
\expandafter\ifx\csname urlstyle\endcsname\relax
  \providecommand{\doi}[1]{doi:\discretionary{}{}{}#1}\else
  \providecommand{\doi}{doi:\discretionary{}{}{}\begingroup
  \urlstyle{rm}\Url}\fi
\providecommand{\eprint}[2][]{\url{#2}}

\bibitem[{Anselin(1995)}]{anselin1995}
Anselin L (1995).
\newblock \enquote{Local Indicators of Spatial Association-{LISA}.}
\newblock \emph{Geographical Analysis}, \textbf{27}(2), 93--115.
\newblock \doi{10.1111/j.1538-4632.1995.tb00338.x}.

\bibitem[{Anselin(2018)}]{anselin2018}
Anselin L (2018).
\newblock \enquote{A Local Indicator of Multivariate Spatial Association:
  Extending Geary's {\emph{c}}.}
\newblock \emph{Geographical Analysis}, \textbf{51}(2), 133--150.
\newblock \doi{10.1111/gean.12164}.

\bibitem[{Baston(2022)}]{exactextractr}
Baston D (2022).
\newblock \emph{{exactextractr}: Fast Extraction from Raster Datasets using
  Polygons}.
\newblock R package version 0.9.1,
  \urlprefix\url{https://CRAN.R-project.org/package=exactextractr}.

\bibitem[{Bivand(2022)}]{bivand2022}
Bivand R (2022).
\newblock \enquote{{R} Packages for Analyzing Spatial Data: A Comparative Case
  Study with Areal Data.}
\newblock \emph{Geographical Analysis}, \textbf{54}(3), 488--518.
\newblock \doi{10.1111/gean.12319}.

\bibitem[{Bivand \emph{et~al.}(2008)Bivand, Pebesma, and
  {Gómez-Rubio}}]{applied2008}
Bivand RS, Pebesma EJ, {Gómez-Rubio} V (2008).
\newblock \emph{Applied Spatial Data Analysis with R}.
\newblock Springer New York, New York.
\newblock \doi{10.1007/978-0-387-78171-6}.

\bibitem[{Bivand and Wong(2018)}]{spdep}
Bivand RS, Wong DWS (2018).
\newblock \enquote{Comparing Implementations of Global and Local Indicators of
  Spatial Association.}
\newblock \emph{TEST}, \textbf{27}(3), 716--748.
\newblock \doi{10.1007/s11749-018-0599-x}.

\bibitem[{Blackard \emph{et~al.}(2008)Blackard, Finco, Helmer, Holden, Hoppus,
  Jacobs, Lister, Moisen, Nelson, and Riemann}]{blackard2008}
Blackard J, Finco M, Helmer E, Holden G, Hoppus M, Jacobs D, Lister A, Moisen
  G, Nelson M, Riemann R (2008).
\newblock \enquote{Mapping {U.S.} Forest Biomass Using Nationwide Forest
  Inventory Data and Moderate Resolution Information.}
\newblock \emph{Remote Sensing of Environment}, \textbf{112}(4), 1658--1677.
\newblock \doi{10.1016/j.rse.2007.08.021}.

\bibitem[{Correndo \emph{et~al.}(2022)Correndo, {Moro Rosso}, Schwalbert,
  Hernandez, Bastos, Nieto, Holzworth, and Ciampitti}]{metrica}
Correndo AA, {Moro Rosso} LH, Schwalbert R, Hernandez C, Bastos LM, Nieto L,
  Holzworth D, Ciampitti IA (2022).
\newblock \emph{{metrica}: Prediction Performance Metrics}.
\newblock R package version 2.0.1,
  \urlprefix\url{https://CRAN.R-project.org/package=metrica}.

\bibitem[{Draper and Yang(1997)}]{draper1997}
Draper NR, Yang Y (1997).
\newblock \enquote{Generalization of the Geometric Mean Functional
  Relationship.}
\newblock \emph{Computational Statistics \& Data Analysis}, \textbf{23}(3),
  355--372.
\newblock \doi{10.1016/s0167-9473(96)00037-0}.

\bibitem[{Fick and Hijmans(2017)}]{worldclim}
Fick SE, Hijmans RJ (2017).
\newblock \enquote{WorldClim 2: New 1{-}km Spatial Resolution Climate Surfaces
  for Global Land Areas.}
\newblock \emph{International Journal of Climatology}, \textbf{37}(12),
  4302--4315.
\newblock \doi{10.1002/joc.5086}.

\bibitem[{Frick \emph{et~al.}(2022)Frick, Chow, Kuhn, Mahoney, Silge, and
  Wickham}]{rsample}
Frick H, Chow F, Kuhn M, Mahoney M, Silge J, Wickham H (2022).
\newblock \emph{{rsample}: General Resampling Infrastructure}.
\newblock R package version 1.1.1,
  \urlprefix\url{https://CRAN.R-project.org/package=rsample}.

\bibitem[{Geary(1954)}]{geary1954}
Geary RC (1954).
\newblock \enquote{The Contiguity Ratio and Statistical Mapping.}
\newblock \emph{The Incorporated Statistician}, \textbf{5}(3), 115.
\newblock \doi{10.2307/2986645}.

\bibitem[{Getis and Ord(1992)}]{getis2010}
Getis A, Ord JK (1992).
\newblock \enquote{The Analysis of Spatial Association by Use of Distance
  Statistics.}
\newblock \emph{Geographical Analysis}, \textbf{24}(3), 189--206.
\newblock \doi{10.1111/j.1538-4632.1992.tb00261.x}.

\bibitem[{Getis and Ord(1996)}]{getis1996local}
Getis A, Ord JK (1996).
\newblock \emph{Spatial Analysis: Modelling in a {GIS} Environment}, chapter
  Local Spatial Statistics: An Overview, pp. 261--277.
\newblock GeoInformation International, Cambridge, UK.

\bibitem[{Gotti and Kuhn(2022)}]{applicable}
Gotti M, Kuhn M (2022).
\newblock \emph{{applicable}: A Compilation of Applicability Domain Methods}.
\newblock R package version 0.1.0,
  \urlprefix\url{https://CRAN.R-project.org/package=applicable}.

\bibitem[{Greenwell and Boehmke(2020)}]{vip}
Greenwell BM, Boehmke BC (2020).
\newblock \enquote{Variable Importance Plots — An Introduction to the {vip}
  Package.}
\newblock \emph{{The R Journal}}, \textbf{12}(1), 343--366.
\newblock \doi{10.32614/RJ-2020-013}.

\bibitem[{Guerry(1833)}]{guerry}
Guerry AM (1833).
\newblock \emph{Essai sur la statistique morale de la France}.
\newblock Crochard, Paris.

\bibitem[{Hijmans(2023)}]{terra}
Hijmans RJ (2023).
\newblock \emph{{terra}: Spatial Data Analysis}.
\newblock R package version 1.7-3,
  \urlprefix\url{https://CRAN.R-project.org/package=terra}.

\bibitem[{Ji and Gallo(2006)}]{ji2006}
Ji L, Gallo K (2006).
\newblock \enquote{An Agreement Coefficient for Image Comparison.}
\newblock \emph{Photogrammetric Engineering \& Remote Sensing}, \textbf{72}(7),
  823--833.
\newblock \doi{10.14358/pers.72.7.823}.

\bibitem[{Johnson \emph{et~al.}(2022)Johnson, Mahoney, Bevilacqua, Stehman,
  Domke, and Beier}]{johnson2022}
Johnson LK, Mahoney MJ, Bevilacqua E, Stehman SV, Domke GM, Beier CM (2022).
\newblock \enquote{Fine-Resolution Landscape-Scale Biomass Mapping Using a
  Spatiotemporal Patchwork of {LiDAR} Coverages.}
\newblock \emph{International Journal of Applied Earth Observation and
  Geoinformation}, \textbf{114}, 103059.
\newblock \doi{10.1016/j.jag.2022.103059}.

\bibitem[{Kuhn(2022{\natexlab{a}})}]{caret}
Kuhn M (2022{\natexlab{a}}).
\newblock \emph{{caret}: Classification and Regression Training}.
\newblock R package version 6.0-93,
  \urlprefix\url{https://CRAN.R-project.org/package=caret}.

\bibitem[{Kuhn(2022{\natexlab{b}})}]{tune}
Kuhn M (2022{\natexlab{b}}).
\newblock \emph{{tune}: Tidy Tuning Tools}.
\newblock R package version 1.0.1,
  \urlprefix\url{https://CRAN.R-project.org/package=tune}.

\bibitem[{Kuhn and Frick(2022)}]{dials}
Kuhn M, Frick H (2022).
\newblock \emph{{dials}: Tools for Creating Tuning Parameter Values}.
\newblock R package version 1.1.0,
  \urlprefix\url{https://CRAN.R-project.org/package=dials}.

\bibitem[{Kuhn and Silge(2022)}]{tidymodels}
Kuhn M, Silge J (2022).
\newblock \emph{Tidy Modeling with {R}}.
\newblock O'Reilly, Sebastopol, CA.

\bibitem[{Kuhn and Vaughan(2022)}]{parsnip}
Kuhn M, Vaughan D (2022).
\newblock \emph{{parsnip}: A Common {API} to Modeling and Analysis Functions}.
\newblock R package version 1.0.3,
  \urlprefix\url{https://CRAN.R-project.org/package=parsnip}.

\bibitem[{Kuhn \emph{et~al.}(2023)Kuhn, Vaughan, and Hvitfeldt}]{yardstick}
Kuhn M, Vaughan D, Hvitfeldt E (2023).
\newblock \emph{{yardstick}: Tidy Characterizations of Model Performance}.
\newblock R package version 1.2.0,
  \urlprefix\url{https://CRAN.R-project.org/package=yardstick}.

\bibitem[{Legates and McCabe(1999)}]{legates1999}
Legates DR, McCabe GJ (1999).
\newblock \enquote{Evaluating the Use of
  {\textquotedblleft}Goodness-of-Fit{\textquotedblright} Measures in Hydrologic
  and Hydroclimatic Model Validation.}
\newblock \emph{Water Resources Research}, \textbf{35}(1), 233--241.
\newblock \doi{10.1029/1998wr900018}.

\bibitem[{Legendre and Fortin(1989)}]{legendre1989}
Legendre P, Fortin M (1989).
\newblock \enquote{Spatial Pattern and Ecological Analysis.}
\newblock \emph{Vegetatio}, \textbf{80}(2), 107--138.
\newblock \doi{10.1007/bf00048036}.

\bibitem[{Leroy \emph{et~al.}(2015)Leroy, Meynard, Bellard, and
  Courchamp}]{virtualspecies}
Leroy B, Meynard CN, Bellard C, Courchamp F (2015).
\newblock \enquote{{virtualspecies}, an {R} Package to Generate Virtual Species
  Distributions.}
\newblock \emph{Ecography}.
\newblock \doi{10.1111/ecog.01388}.

\bibitem[{Li and Anselin(2023)}]{rgeoda}
Li X, Anselin L (2023).
\newblock \emph{{rgeoda}: {R} Library for Spatial Data Analysis}.
\newblock R package version 0.0.10-2,
  \urlprefix\url{https://CRAN.R-project.org/package=rgeoda}.

\bibitem[{Mahoney and Silge(2023)}]{spatialsample}
Mahoney M, Silge J (2023).
\newblock \emph{{spatialsample}: Spatial Resampling Infrastructure}.
\newblock R package version 0.3.0,
  \urlprefix\url{https://CRAN.R-project.org/package=spatialsample}.

\bibitem[{Meyer \emph{et~al.}(2023)Meyer, {Milà}, and Ludwig}]{CAST}
Meyer H, {Milà} C, Ludwig M (2023).
\newblock \emph{{CAST}: '{caret}' Applications for Spatial-Temporal Models}.
\newblock R package version 0.7.1,
  \urlprefix\url{https://CRAN.R-project.org/package=CAST}.

\bibitem[{Meyer and Pebesma(2021)}]{meyer2021}
Meyer H, Pebesma E (2021).
\newblock \enquote{Predicting into Unknown Space? Estimating the Area of
  Applicability of Spatial Prediction Models.}
\newblock \emph{Methods in Ecology and Evolution}, \textbf{12}(9), 1620--1633.
\newblock \doi{10.1111/2041-210x.13650}.

\bibitem[{Moran(1950)}]{moran1950}
Moran PAP (1950).
\newblock \enquote{Notes on Continuous Stochastic Phenomena.}
\newblock \emph{Biometrika}, \textbf{37}(1/2), 17.
\newblock \doi{10.2307/2332142}.

\bibitem[{{Müller} and Wickham(2022)}]{tibble}
{Müller} K, Wickham H (2022).
\newblock \emph{{tibble}: Simple Data Frames}.
\newblock R package version 3.1.8,
  \urlprefix\url{https://CRAN.R-project.org/package=tibble}.

\bibitem[{Nelson \emph{et~al.}(2009)Nelson, McRoberts, Holden, and
  Bauer}]{nelson2009}
Nelson MD, McRoberts RE, Holden GR, Bauer ME (2009).
\newblock \enquote{Effects of Satellite Image Spatial Aggregation and
  Resolution on Estimates of Forest Land Area.}
\newblock \emph{International Journal of Remote Sensing}, \textbf{30}(8),
  1913--1940.
\newblock \doi{10.1080/01431160802545631}.

\bibitem[{Netzeva \emph{et~al.}(2005)Netzeva, Worth, Aldenberg, Benigni,
  Cronin, Gramatica, Jaworska, Kahn, Klopman, Marchant, Myatt,
  Nikolova-Jeliazkova, Patlewicz, Perkins, Roberts, Schultz, Stanton, van~de
  Sandt, Tong, Veith, and Yang}]{netzeva2005}
Netzeva TI, Worth AP, Aldenberg T, Benigni R, Cronin MT, Gramatica P, Jaworska
  JS, Kahn S, Klopman G, Marchant CA, Myatt G, Nikolova-Jeliazkova N, Patlewicz
  GY, Perkins R, Roberts DW, Schultz TW, Stanton DT, van~de Sandt JJ, Tong W,
  Veith G, Yang C (2005).
\newblock \enquote{Current Status of Methods for Defining the Applicability
  Domain of (Quantitative) Structure-Activity Relationships.}
\newblock \emph{Alternatives to Laboratory Animals}, \textbf{33}(2), 155--173.
\newblock \doi{10.1177/026119290503300209}.

\bibitem[{Ord and Getis(1995)}]{ord2010}
Ord JK, Getis A (1995).
\newblock \enquote{Local Spatial Autocorrelation Statistics: Distributional
  Issues and an Application.}
\newblock \emph{Geographical Analysis}, \textbf{27}(4), 286--306.
\newblock \doi{10.1111/j.1538-4632.1995.tb00912.x}.

\bibitem[{Pebesma(2018)}]{sf}
Pebesma E (2018).
\newblock \enquote{Simple Features for {R}: Standardized Support for Spatial
  Vector Data.}
\newblock \emph{{The R Journal}}, \textbf{10}(1), 439--446.
\newblock \doi{10.32614/RJ-2018-009}.

\bibitem[{Pedersen(2022)}]{patchwork}
Pedersen TL (2022).
\newblock \emph{patchwork: The Composer of Plots}.
\newblock R package version 1.1.2,
  \urlprefix\url{https://CRAN.R-project.org/package=patchwork}.

\bibitem[{Reich and Barai(1999)}]{reich1999}
Reich Y, Barai S (1999).
\newblock \enquote{Evaluating Machine Learning Models for Engineering
  Problems.}
\newblock \emph{Artificial Intelligence in Engineering}, \textbf{13}(3),
  257--272.
\newblock \doi{10.1016/s0954-1810(98)00021-1}.

\bibitem[{Riemann \emph{et~al.}(2010)Riemann, Wilson, Lister, and
  Parks}]{riemann2010}
Riemann R, Wilson BT, Lister A, Parks S (2010).
\newblock \enquote{An Effective Assessment Protocol for Continuous Geospatial
  Datasets of Forest Characteristics Using {USFS} Forest Inventory and Analysis
  ({FIA}) Data.}
\newblock \emph{Remote Sensing of Environment}, \textbf{114}(10), 2337--2352.
\newblock \doi{10.1016/j.rse.2010.05.010}.

\bibitem[{Roehm \emph{et~al.}(2012)Roehm, Tiarks, Koschke, and
  Maalej}]{roehm2012}
Roehm T, Tiarks R, Koschke R, Maalej W (2012).
\newblock \enquote{How do Professional Developers Comprehend Software?}
\newblock \emph{2012 34th International Conference on Software Engineering
  (ICSE)}.
\newblock \doi{10.1109/icse.2012.6227188}.
\newblock \urlprefix\url{http://dx.doi.org/10.1109/ICSE.2012.6227188}.

\bibitem[{van~den Hoogen \emph{et~al.}(2019)van~den Hoogen, Geisen, Routh,
  Ferris, Traunspurger, Wardle, de~Goede, Adams, Ahmad, Andriuzzi, Bardgett,
  Bonkowski, Campos-Herrera, Cares, Caruso, de~Brito~Caixeta, Chen, Costa,
  Creamer, Mauro~da Cunha~Castro, Dam, Djigal, Escuer, Griffiths, {Gutiérrez},
  Hohberg, Kalinkina, Kardol, Kergunteuil, Korthals, Krashevska, Kudrin, Li,
  Liang, Magilton, Marais, {Martín}, Matveeva, Mayad, Mulder, Mullin, Neilson,
  Nguyen, Nielsen, Okada, Rius, Pan, Peneva, Pellissier, Carlos Pereira~da
  Silva, Pitteloud, Powers, Powers, Quist, Rasmann, Moreno, Scheu, {Setälä},
  Sushchuk, Tiunov, Trap, van~der Putten, {Vestergård}, Villenave,
  Waeyenberge, Wall, Wilschut, Wright, Yang, and Crowther}]{vandenhoogen2019}
van~den Hoogen J, Geisen S, Routh D, Ferris H, Traunspurger W, Wardle DA,
  de~Goede RGM, Adams BJ, Ahmad W, Andriuzzi WS, Bardgett RD, Bonkowski M,
  Campos-Herrera R, Cares JE, Caruso T, de~Brito~Caixeta L, Chen X, Costa SR,
  Creamer R, Mauro~da Cunha~Castro J, Dam M, Djigal D, Escuer M, Griffiths BS,
  {Gutiérrez} C, Hohberg K, Kalinkina D, Kardol P, Kergunteuil A, Korthals G,
  Krashevska V, Kudrin AA, Li Q, Liang W, Magilton M, Marais M, {Martín} J,
  Matveeva E, Mayad EH, Mulder C, Mullin P, Neilson R, Nguyen TAD, Nielsen UN,
  Okada H, Rius JEP, Pan K, Peneva V, Pellissier L, Carlos Pereira~da Silva J,
  Pitteloud C, Powers TO, Powers K, Quist CW, Rasmann S, Moreno S, Scheu S,
  {Setälä} H, Sushchuk A, Tiunov AV, Trap J, van~der Putten W, {Vestergård}
  M, Villenave C, Waeyenberge L, Wall DH, Wilschut R, Wright DG, Yang Ji,
  Crowther TW (2019).
\newblock \enquote{Soil Nematode Abundance and Functional Group Composition at
  a Global Scale.}
\newblock \emph{Nature}, \textbf{572}(7768), 194--198.
\newblock \doi{10.1038/s41586-019-1418-6}.

\bibitem[{Vaughan and Couch(2022)}]{workflows}
Vaughan D, Couch S (2022).
\newblock \emph{{workflows}: Modeling Workflows}.
\newblock R package version 1.1.2,
  \urlprefix\url{https://CRAN.R-project.org/package=workflows}.

\bibitem[{Wickham(2016)}]{ggplot2}
Wickham H (2016).
\newblock \emph{ggplot2: Elegant Graphics for Data Analysis}.
\newblock Springer-Verlag New York.
\newblock ISBN 978-3-319-24277-4.
\newblock \urlprefix\url{https://ggplot2.tidyverse.org}.

\bibitem[{Wickham \emph{et~al.}(2023{\natexlab{a}})Wickham, François, Henry,
  Müller, and Vaughan}]{dplyr}
Wickham H, François R, Henry L, Müller K, Vaughan D (2023{\natexlab{a}}).
\newblock \emph{{dplyr}: A Grammar of Data Manipulation}.
\newblock R package version 1.1.0,
  \urlprefix\url{https://CRAN.R-project.org/package=dplyr}.

\bibitem[{Wickham \emph{et~al.}(2023{\natexlab{b}})Wickham, Vaughan, and
  Girlich}]{tidyr}
Wickham H, Vaughan D, Girlich M (2023{\natexlab{b}}).
\newblock \emph{{tidyr}: Tidy Messy Data}.
\newblock R package version 1.3.0,
  \urlprefix\url{https://CRAN.R-project.org/package=tidyr}.

\bibitem[{Willmott(1981)}]{willmott1981}
Willmott CJ (1981).
\newblock \enquote{On the Validation of Models.}
\newblock \emph{Physical Geography}, \textbf{2}(2), 184--194.
\newblock \doi{10.1080/02723646.1981.10642213}.

\bibitem[{Willmott(1982)}]{willmott1982}
Willmott CJ (1982).
\newblock \enquote{Some Comments on the Evaluation of Model Performance.}
\newblock \emph{Bulletin of the American Meteorological Society},
  \textbf{63}(11), 1309--1313.
\newblock \doi{10.1175/1520-0477(1982)063<1309:scoteo>2.0.co;2}.

\bibitem[{Willmott \emph{et~al.}(1985)Willmott, Ackleson, Davis, Feddema,
  Klink, Legates, {O'Donnell}, and Rowe}]{willmott1985}
Willmott CJ, Ackleson SG, Davis RE, Feddema JJ, Klink KM, Legates DR,
  {O'Donnell} J, Rowe CM (1985).
\newblock \enquote{Statistics for the Evaluation and Comparison of Models.}
\newblock \emph{Journal of Geophysical Research}, \textbf{90}(C5), 8995.
\newblock \doi{10.1029/jc090ic05p08995}.

\bibitem[{Willmott \emph{et~al.}(2011)Willmott, Robeson, and
  Matsuura}]{willmott2011}
Willmott CJ, Robeson SM, Matsuura K (2011).
\newblock \enquote{A Refined Index of Model Performance.}
\newblock \emph{International Journal of Climatology}, \textbf{32}(13),
  2088--2094.
\newblock \doi{10.1002/joc.2419}.

\bibitem[{Willmott and Wicks(1980)}]{willmott1980}
Willmott CJ, Wicks DE (1980).
\newblock \enquote{An Empirical Method for the Spatial Interpolation of Monthly
  Precipitation Within California.}
\newblock \emph{Physical Geography}, \textbf{1}(1), 59--73.
\newblock \doi{10.1080/02723646.1980.10642189}.

\bibitem[{{Zambrano-Bigiarini}(2020)}]{hydroGOF}
{Zambrano-Bigiarini} M (2020).
\newblock \emph{{hydroGOF}: Goodness-of-Fit Functions for Comparison of
  Simulated and Observed Hydrological Time Series}.
\newblock \doi{10.5281/zenodo.839854}.
\newblock R package version 0.4-0,
  \urlprefix\url{https://github.com/hzambran/hydroGOF}.

\end{thebibliography}

\end{document}